\documentclass[twocolumn,aps,superscriptaddress,prL,floatfix,showpacs,longbibliography]{revtex4-2}

\makeatletter
\newcommand*{\rom}[1]{\expandafter\@slowromancap\romannumeral #1@} 
\makeatother
\usepackage{gensymb} 
\usepackage{graphicx}
\usepackage{times}
\usepackage{dcolumn}
\usepackage{hyperref}
\usepackage{color}
\usepackage{float}
\usepackage{amsmath, amssymb}
\usepackage{courier}
\usepackage{xr-hyper}
\usepackage{ulem}
\usepackage{textcomp,mathcomp}

\newcommand\LNO[0]{La$_3$Ni$_2$O$_{7-\delta}$}
\newcommand\LNOO[0]{La$_3$Ni$_2$O$_{6.92}$}
\newcommand\musr[0]{$\mu^+$SR}

\hypersetup{colorlinks = true, linkcolor = blue, anchorcolor =blue, citecolor = blue, filecolor = blue, urlcolor = blue,
	pdfauthor=author}
\begingroup

\begin{document}
	
	
	\title{Evidence of spin density waves in \LNO}

	\author{Kaiwen Chen}
	\altaffiliation{These authors contributed equally to this work.}
	\affiliation{State Key Laboratory of Surface Physics, Department of Physics, Fudan University, Shanghai 200438, China}
	\author{Xiangqi Liu}
	\altaffiliation{These authors contributed equally to this work.}
	\affiliation{School of Physical Science and Technology, ShanghaiTech University, Shanghai 201210, China}
	\author{Jiachen Jiao}
	\altaffiliation{These authors contributed equally to this work.}
	\affiliation{State Key Laboratory of Surface Physics, Department of Physics, Fudan University, Shanghai 200438, China}
	\author{Muyuan Zou}
	\affiliation{State Key Laboratory of Surface Physics, Department of Physics, Fudan University, Shanghai 200438, China}
	\author{Chengyu Jiang}
	\affiliation{State Key Laboratory of Surface Physics, Department of Physics, Fudan University, Shanghai 200438, China}
	\author{Xin Li}
	\affiliation{State Key Laboratory of Surface Physics, Department of Physics, Fudan University, Shanghai 200438, China}
	\author{Yixuan Luo}
	\affiliation{School of Physical Science and Technology, ShanghaiTech University, Shanghai 201210, China}
	\author{Qiong Wu}
	\affiliation{State Key Laboratory of Surface Physics, Department of Physics, Fudan University, Shanghai 200438, China}
	\author{Ningyuan Zhang}
	\affiliation{State Key Laboratory of Surface Physics, Department of Physics, Fudan University, Shanghai 200438, China}
	\author{Yanfeng Guo}
	\affiliation{School of Physical Science and Technology, ShanghaiTech University, Shanghai 201210, China}
	\affiliation{ShanghaiTech Laboratory for Topological Physics, Shanghai 201210, China}
	\author{Lei Shu}
	\email{leishu@fudan.edu.cn}
	\affiliation{State Key Laboratory of Surface Physics, Department of Physics, Fudan University, Shanghai 200438, China}
	\affiliation{Shanghai Research Center for Quantum Sciences, Shanghai 201315, China}

	
	\begin{abstract} 
		The recently discovered superconductivity with critical temperature $T_c$ up to 80 K in the double-layer Nickelate \LNO\ under pressure has drawn great attention. Here we report the positive muon spin relaxation (\musr) study of polycrystalline \LNOO\ under ambient pressure. Zero-field \musr\ experiments reveal the existence of magnetic order in \LNOO\ with $T_{N}=154\ \rm{K}$. The weak transverse field $\mu^+$SR measurements confirms the bulk nature of magnetism. In addition, a small quantity of oxygen deficiencies can greatly broaden the internal magnetic field distribution sensed by muons.
	\end{abstract}
	
	\maketitle
	\date{}
	
	\textit{Introduction.}---The recent observation of the sign of superconductivity with $T_c\approx$ 80 K in La$_3$Ni$_2$O$_{7}$ single crystals under pressure has attracted significant attention~\cite{Sun2023}. Subsequent high-pressure measurements on single crystal~\cite{Hou2023,Zhang2023} and powder samples~\cite{Wang2023} detected the zero-resistance under more hydrostatic pressure conditions, representing the discovery of a brand-new high-temperature superconductor. This is another discovery of superconductivity with critical temperature higher than the boiling point of liquid nitrogen, after copper oxides~\cite{Bednorz1986,Wu1987,Maeda1988,Sheng1988}, iron-based superconductors~\cite{Ge2015} and hydrides under high pressure~\cite{Eremets2022}. Recent high-pressure alternating-current magnetic susceptibility measurements further confirmed bulk superconductivity in this compound~\cite{Jing2024}. \LNO\ therefore provides an excitingly new opportunity to investigate the pairing mechanism of high-temperature superconductivity. A lot of theoretical works quickly followed~\cite{Luo2023,Qin2023,Shilenko2023,Yang2023_theory,YYF2023,Cao2023,Chen2023,Christiansson2023,Jiang2023,LaBollita2023,Shen2023,Zhang2023_theory,Tian2023,Sakakibara2023,Gu2023,Zhang2023structural}, some of which proposed $s\pm$-pairing superconductivity triggered by spin fluctuations under high pressure~\cite{Yang2023_theory,Zhang2023_theory,Tian2023,Sakakibara2023,Gu2023,Zhang2023structural}. On the other hand, due to the necessity of high pressure to induce superconductivity, experimental progress on pairing mechanism is quite limited~\cite{Wang2023,Hou2023,Zhang2023,Yang2023,Zhou2023,Yuan2023,WHH2023}. 
	
	Recent physical property measurements on single crystal La$_3$Ni$_2$O$_{7}$ revealed a density-wave like transition near 153 K at ambient pressure~\cite{Liu2023}. The onset temperature of the anomaly will be suppressed by pressure, followed by the emergence of superconductivity under higher pressure~\cite{Hou2023,Zhang2023}. 
	However, the nature of such density-wave like transition is still unclear. Resonant inelastic x-ray scattering (RIXS)~\cite{RIXS} and nuclear magnetic resonant (NMR)~\cite{NMR2024} suggest the spin-density-wave below 150 K, while no magnetic order is identified by neutron scattering measurement down to 10 K~\cite{NPD2024}. A rich interplay between magnetic order and superconductivity is the key character of unconventional superconductors, where superconductivity often appears near the border of magnetic order, e.g., copper oxides~\cite{Niedermayer1998,Julien2003,Sanna2004}, iron pnictides~\cite{Paglione2010,Dai2015} and heavy fermion superconductors~\cite{Kenzelmann2008}. Therefore, elucidating the magnetic ground state of \LNO\ is the core issue on high temperature superconducting mechanism.
	
	Here, we report the \musr\ measurements on polycrystalline \LNO\ to clarify the magnetic ground state of \LNO\ at ambient pressure. Positive muon spin relaxation/rotation (\musr) is an unmatched technique for detecting magnetism or spin dynamics~\cite{Blundell1999,Yaouanc2011MuonSR,Adrian2022}. 100\% polarized muons are implanted into the sample, which work as sensitive local spin probe. Zero-field (ZF)-\musr\ measurements reveal the commensurate magnetic order in \LNO. Weak transverse field (wTF)-\musr\ experiments confirm the bulk nature of magnetism.  On the other hand, the present \musr\ measurement observes inhomogenous internal magnetic field in powder \LNO, which further highlights the influence of oxygen vacancies on magnetism in \LNO.
	
	
	
	\textit{Experimental Details.}---Polycrystalline \LNO\ were synthesized through the solid-state reaction. High-purity La$_2$O$_3$ (99.999\%, Aladdin) and NiO (99.99\%, Macklin) were mixed with a molar ratio of 3:4.01. The slight excess of NiO was used to compensate the loss of volatilization. The ground mixtures were sintered at 1100 ℃ in the air for 50 h. The reactants were reground and sintered again for 3 times to get completely reacted and hence homogeneous polycrystalline samples. Powder x-ray diffraction (PXRD) patterns were obtained by using a Bruker D8 advanced x-ray diffraction spectrometer ($\lambda$ = 1.5418 \r{A})
	at room temperature. The XRD Rietveld refinement was conducted with \texttt{Fullprof} software~\cite{FULLPROF}. The oxygen content was obtained by the thermogravimetric analysis (TGA) technique in METTLER TOLEDO TGA/DSC3+, using a 10\% H$_2$/Ar gas flow of 50 mL/min and heating up from room temperature to 950 $\tccelsius$ with a 7.5 $\tccelsius$/min rate. The magnetization of \LNO\  was measured in a superconducting quantum interference device magnetometer (Quantum Design magnetic property measurement system).
	The temperature-dependent susceptibility between 2 and 300 K was measured under a magnetic field of 0.4 T in both zero-field cooled (ZFC) and field-cooled (FC) procedures. Temperature dependence of resistivity $\rho(T)$ was measured using standard four-probe method with a physical property measurement system (PPMS). Powder samples were pressed and cut into rectangle. Four annealed silver wires were glued on the surface of the sample with conductive silver adhesives. The resistivities were measured between 2 K and 300 K by cooling.
	
	The ZF, LF and wTF-\musr\ experiments were carried out on M15 and M20 spectrometers at TRIUMF, Vancouver. About 350 mg of the powder sample was pressed into rounds with the diameter of about 1.2 mm. The sample of M15 was held on a high-purity silver plate with diluted GE varnish and loaded in a top-loading dilution refrigerator with a base temperature of 35 mK. The sample measured at M20 was mounted on a hollow square copper frame with thin silver tape and was free from additional background signal. The \musr\ measurement at M15 was carried out down to 35 mK in zero field and in a longitudinal field at 0.1 T. The \musr\ measurement at M20 was carried  out between 3.5 K and 290 K in zero field and in a transverse field at 30 Oe. The temperature was carefully controlled to ensure that the standard deviation of the temperature during each measurement was less than 0.1 K. The \musr\  data were analyzed using the \texttt{MUSRFIT} software package~\cite{MUSRFIT}. 
	
\begin{figure}[ht]
		\includegraphics[width=0.5\textwidth]{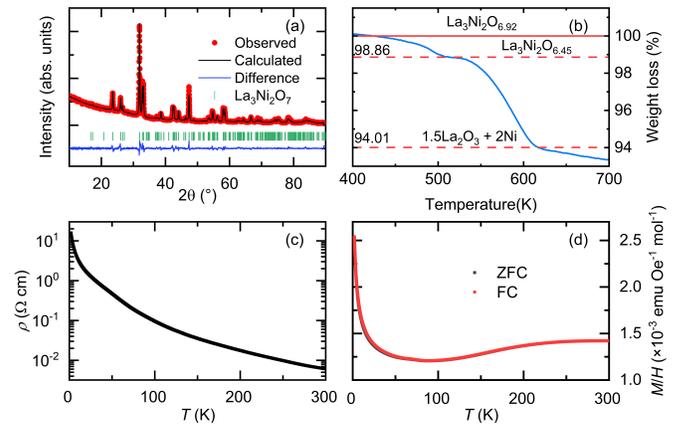}
		\caption{\label{fig:Fig1}(a) Rietveld refinement of powder-pattern X-ray diffraction data; (b) Thermogravimetric curves for \LNO\ in 10\% H$_2$/Ar from 400 \textcelsius\ to 700 \textcelsius; (c) temperature dependence of the resistivity $\rho(T)$ of polycrystalline \LNOO; (d) temperature dependence of magnetic susceptibility $\chi(T)$ between 2 K and 300 K with $\mu_0H=0.4$ T. No anomaly is observed in $\rho(T)$ and $\chi(T)$ between 100 and 200 K.}
	\end{figure}

	\textit{Physical Properties.}---Fig.~\ref{fig:Fig1}(a) shows the x-ray diffraction pattern of polycrystalline \LNO. No impurity phase is detected by the x-ray diffraction. The Rietveld refinement confirms an orthorhombic structure (space group \textit{Amam}, no. 63) with $\textit{a}=5.4018(5)$ $\rm{\r{A}}, \textit{b}=5.4557(7)\rm{\r{A}}$, $\textit{c}=20.537(2)\ \rm{\r{A}}$ and $\chi^2$ = 4.67. The obtained parameter structures are consistent with the results reported in Ref.~\cite{Wang2023}.
	The oxygen stoichiometry is determined by the TGA measurement with a 10\% H$_2$/Ar flow. As shown in Fig.~\ref{fig:Fig1}(b), the reduction of polycrystalline \LNO\ occurs in two steps. The oxygen stoichiometry of the initial and the intermediate phase is determined by calculating the weight loss between the phases, which results $\rm{La_3Ni_2O_{6.92(1)}}$ and $\rm{La_3Ni_2O_{6.45(1)}}$, respectively. The final formation of $\rm{La_2O_3}$ and $\rm{Ni}$ is confirmed by powder XRD measurement. Combining the results of PXRD and TGA measurements, we use \LNOO\ to refer to our sample in the following text.
	
	Temperature dependence of resistivity $\rho(T)$ of polycrystalline \LNOO\ is plotted in Fig.~\ref{fig:Fig1}(c). The value of resistivity is relatively large compared to reported results~\cite{Zhang1994,Tian2023,Wang2023}. It should be noted that the measured \LNO\ powder was pressed into pellet without sintering before the resistivity measurement. The resulted powder pellet is relatively loose and is easily disintegrated, this could greatly increase the absolute value of resistivity. Despite the large value, $\rho(T)$ shows a negative temperature coefficient and no anomaly is detected around 150~K.
	
	Fig.\ref{fig:Fig1}(d) displays the temperature dependence of dc magnetic susceptibility $\chi(T)$, which was measured under the magnetic field of $\mu_0H=0.4$ T with both zero-field cooling and field cooling setup. The temperature dependence and magnitude of $\chi(T)$ are consistent with previous results~\cite{Zhang2023,Hou2023,Liu2023} and there is no sign of any magnetic phase transition or spin freezing behavior down to 2 K. The moderate upturn of $\chi(T)$ at low temperature may due to the localized moments induced by randomly distributed oxygen vacancies~\cite{Liu_PRL2023}.
	
\begin{figure}[ht]
		\begin{center}
			\includegraphics[width=0.5\textwidth]{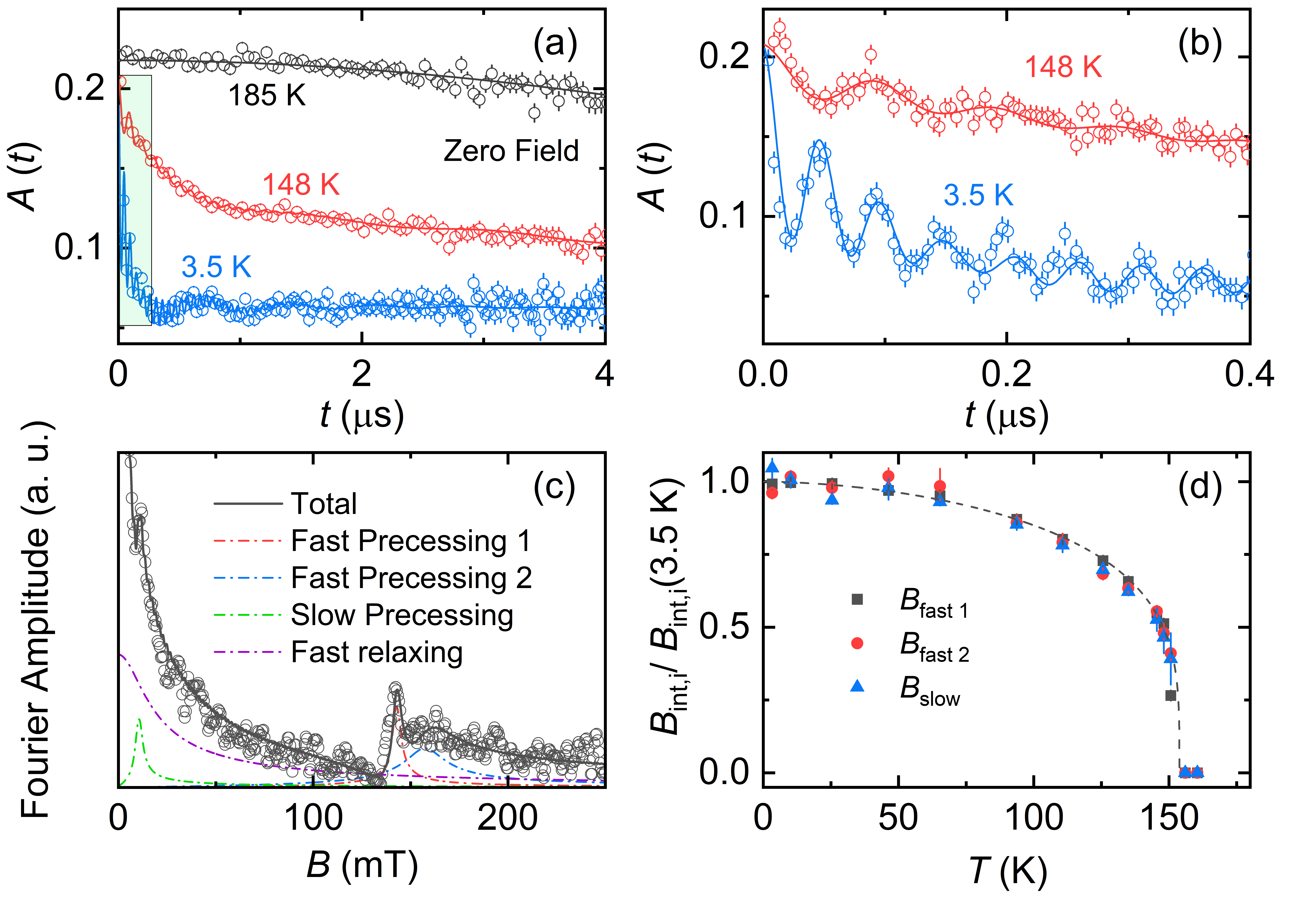}
			\caption{\label{fig:Fig2}(a) Temperature dependence of ZF-\musr\ spectra of \LNOO\ to 4 $\rm{\mu s}$. (b) Early time spectra before 0.4 $\rm{\mu}$s, which shows the zoomed-in area of the green shadow in (a). The solid curves are the fits with Eq.(\ref{eq:2}). (c) The Fourier transform diagram of ZF-\musr\ data measured at 3.5 K. The colored dash lines represent the contribution of four magnetic components. (d) Temperature dependence of magnetic order parameter $B_{\rm{int,i}}$ deduced from the fitting with Eq.(\ref{eq:1}). The black dashed line exhibits the fitting with Eq.~(\ref{eq:2})}
		\end{center}
	\end{figure}

\begin{figure}[ht]
		\begin{center}
			\includegraphics[width=0.48\textwidth]{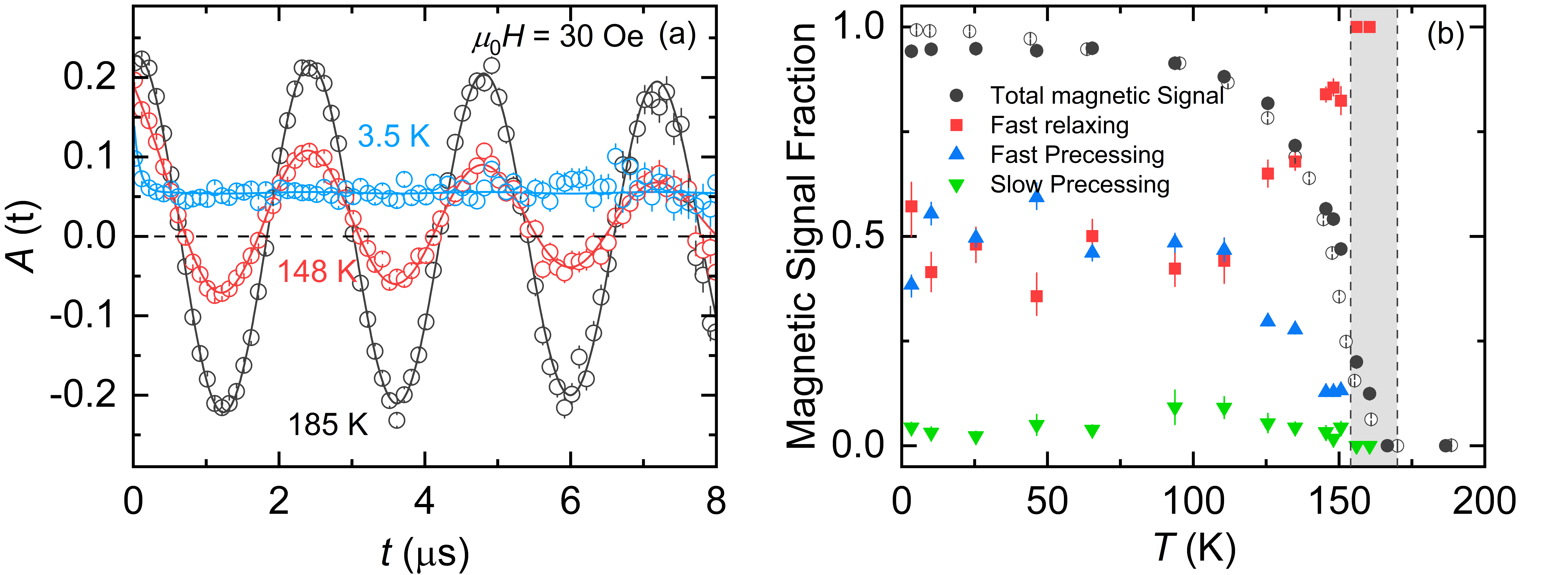}
			\caption{\label{fig:Fig3}(a) Representative wTF-\musr\ spectra of \LNOO\ at different temperatures. The solid curves are the fits of data with Eq.(\ref{eq:2}). (b) Temperature dependence of $f_i$ and $f_m$ extracted from ZF (filled dots) and wTF (hollow circle dots) fitting, respectively. $f_{\rm{fast1}}/f_{\rm{fast2}}=0.73$ is fixed during the fitting and is not shown separately. $f_m$ extracted by Eq.~(\ref{eq:1}) and Eq.~(\ref{eq:3}) are consistent at all temperatures.  The grey shaded area marks the temperature range for the possible spin fluctuation above $T_N$~\cite{NMR2024}.}
		\end{center}
	\end{figure}

	\textit{Zero-field muon spin relaxation.}---ZF-\musr\ is very sensitive to any local magnetic order or magnetic fluctuations~\cite{Hayano1979,Amato1997,Zhang2015}.
	Three representative muon relaxation spectra are shown in Fig.~\ref{fig:Fig2}(a). Above 160 K, the muon spin relaxation spectrum can be described with a simple Gaussian-Kubo-Toyabe function, which describes the muon depolarization due to randomly oriented nuclear dipole moments~\cite{Hayano1979}. An exponentially decaying term appears when cooling the temperature lower than 170 K. However, no oscillation can be identified in the early time spectrum. The oscillating terms appear in the early time window when further cooling down to 150 K (before 0.4 $\mu$s, shown in Fig.~\ref{fig:Fig2}(b)). The frequency and the amplitude of the oscillating component increase as cooling and reach saturation at low temperature.  Fig.~\ref{fig:Fig2}(c) shows the Fourier transform of the data measured at 3.5 K. The Fourier signal is featured with two narrow peaks at 10 mT , 140 mT and a much broader distribution near 160 mT. The ZF-\musr\ spectrum is therefore described with the following formula:
	\begin{multline}
		\label{eq:1}
		A_{\rm{ZF}}/A_0=(1-f_m)G_{\rm{KT}}(\sigma_{\rm{ZF}},t)\\
		+f_m[(1-f_L)\sum_{i=1}f_i\cos(\gamma_{\rm{\mu}}B_{\rm{int,i}}t)e^{-\lambda_it}+f_Le^{{-\lambda_L}t}
		]
	\end{multline}
	$A_0$ is initial asymmetry parameter. The first term describes the nonmagnetic order contribution with volume fraction $1-f_m$. $G_{\rm{KT}}$ is the Gaussian Kubo-Toyabe function described above. $\sigma_{\rm{ZF}}=\delta B_{G}\gamma_{\rm{\mu}}$ describes the distribution width $\delta B_G$ of internal field. $\gamma_{\mu}=851.616\ \rm{MHz/T}$ is the muon gyromagnetic ratio~\cite{Hayano1979}. $\sigma_{\rm{ZF}}$ is found to be temperature independent and is fixed to the average value 0.08 $\rm{\mu} s^{-1}$ during the fitting. The second term describes the magnetic contribution. The non-oscillating term $f_L$ accounts for the muon components with spin aligning along the direction of local magnetic field. $f_L=0.263$ is determined by the long time spectrum (8 $\rm{\mu s}$) of both ZF and wTF-\musr\ measurements. This value is about 80\% of the expected value 1/3 in powder sample, which is most likely due to the preferred orientation induced during pressing~\cite{Sugiyama2009}. $\lambda_{L}$ is about 0.003 $\rm{\mu s^{-1}}$ and shows no temperature dependence. $f_i, \lambda_i$ and $B_{\rm{int,i}}$ are the relative volume fraction, relaxation rate and the magnitude of internal field of the $i$-th magnetic component, respectively.
	The best description of the data requires four components, three oscillating (fast/slow precessing) and one non-oscillating (fast relaxing), in the summation notation of Eq.~(\ref{eq:1}). The fitting using Eq.~(\ref{eq:1}) is shown in Fig.~\ref{fig:Fig2}. The Fourier transform diagram of the four magnetic components is shown in Fig.~\ref{fig:Fig2}(c) with colored dashed lines. $f_i$, $f_m$ and $\lambda_i$ are shown in Fig.~\ref{fig:Fig3}(b) and Fig.~S1~\cite{Supmat}, respectively.
	The weight ration of two fast precessing signals is temperature independent and is fixed to the averaged value between 3.5 K and 93 K during the fitting~\cite{Supmat}. The magnetic fields of three precessing signals exhibit consistent temperature dependence, indicating that there are multiple magnetically inequivalent muon stopping sites in \LNOO. The oscillations can be fitted with sum of cosine functions, which indicates a commensurate magnetic order. This is consistent with the magnon dispersion revealed by recent RIXS experiments~\cite{RIXS}. The reduced temperature dependence of internal magnetic fields is plotted in Fig.~\ref{fig:Fig2}(d).  The temperature dependence of internal field is further fitted with a phenomenological function~\cite{Sugiyama2009}:
	\begin{eqnarray}
		\label{eq:2}
		B_{\rm{int}}(T)=B_{\rm{int}}(0)[1-(T/T_N)^{\alpha}]^{\beta}.
	\end{eqnarray}
	The critical exponents are sensitive to the asymptotic behavior of $B_{\rm{int}}(T)$ approaching $T_N$. The fit is therefore applied to $B_{\rm{fast 1}}$, which is the most recognizable precessing signal. The fitting results in exponent $\alpha=1.7(2), \beta=0.26(2)$, $T_N=154(1)\ \rm{K}$, $B(0)=144(1)\ \rm{mT}$. The fitted critical exponent $\beta$ is smaller than $\beta\approx0.365$ expected by 3D-Heisenberg model~\cite{Critical_exponent}, indicating the reduced dimensionality of the magnetic structure~\cite{Bramwell_1993,RIXS}.
	The fitted $T_N$ is smaller than the magnetic transition starting point near 170 K, which is revealed by magnetic volume fraction $f_m$ extracted from both ZF and wTF-\musr\ measurement (see Fig.~\ref{fig:Fig3}(b)). On the one hand, it is difficult to extract reasonable oscillating frequencies when approaching critical temperature, which is above 150 K in our case. This can lead to an uncertainty of fitted $T_N$. Nevertheless, it is hard to explain a  difference of $\sim$10~K. On the other hand, this may be due to the spin fluctuation over $T_N$ reported in Ref.~\cite{NMR2024}. The \musr\ probe magnetic order on the shorter time scale compared to NMR, therefore senses a lower stating temperature. Unfortunately, this point cannot be judged based on the present work and requires further study. The dynamic nature of the magnetism below $T_N$ is investigated by the longitudinal field (LF) \musr\ measured at 5 K. The muon spin relaxation is completely suppressed with a longitudinal field of 0.1 T~\cite{Supmat}, which is comparable to the magnitude of internal field in the magnetic ordered state. This indicates that the internal field is static or quasi-static compared to the time window of \musr\ technique at low temperature.
	
	\textit{Weak-transverse-field muon spin relaxation.}---The wTF-\musr\ spectra are measured to determine the magnetic volume fraction of \LNOO. In wTF-\musr\ experiment, an external magnetic field perpendicular to the initial muon polarization direction, with a magnitude ($\mu_0H=3\ \rm{mT}$) much smaller than the internal magnetic field ($\mu_0H_{\rm{\rm{int}}}\approx140\ \rm{mT}$), is applied to the sample. The external field is relatively weak so it will not change the magnetic field distribution in the magnetic phase. 
	The muons stopping in the magnetic order phase experience a broader field distribution, and the muon spins depolarize rapidly during the precession. On the other hand, the muons stopping in the non-ordered phase only precess at a frequency corresponding to the external field. Therefore, the relative asymmetry values provide us with the information of the ordered volume fraction of the sample.
	
	The wTF-\musr\ spectra under several temperatures are plotted in Fig.~\ref{fig:Fig3}. With cooling down, the amplitude of the oscillating signal decreases and an exponential-like decaying appears. The oscillating signal disappears at low temperature, indicating the bulk magnetism in our sample. The wTF-\musr\ spectra can be described with the following functional form:
	\begin{multline}
		\label{eq:3}
		A_{\rm{TF}}(t)/A_0=f_m[(1-f_L)e^{-\lambda_{\rm{TF}}t}+f_Le^{-\lambda_{L}t}]\\
		+(1-f_m)\cos(\gamma_{\mu}B_{\rm{ext}}+\phi)e^{\frac{-\sigma^2t^2}{2}}
	\end{multline}
		The second term describes the muons stopping in the non-magnetically ordered phase, which are precessing at frequency $\gamma_{\rm{\mu}}B_{\rm{ext}}$. The Gaussian relaxation term accounts for the field broadening, which is temperature independent and is fixed to 0.08 $\rm{\mu s^{-1}}$ during the fitting. $\phi$ is the initial phase of the implanted  muons. The first term accounts for the muon signal of the magnetic phase. $f_L$ is the longitudinal fraction, which is same as the one described in ZF measurement. Temperature dependence of the magnetically ordered volume fraction $f_m$ is plotted in Fig.~\ref{fig:Fig3}(b) with solid circles. $f_m$ reaches 50\% at 148 K, and 90\% below 93 K, indicating a bulk magnetism in polycrystalline \LNOO. The breadth of magnetic transition indicates that the magnetic field sensed by muons is inhomogenous. This can be due to the existence of oxygen deficiencies discussed below or a short magnetic coherence length~\cite{NPD2024}.

\begin{figure}[ht]
		\begin{center}
			\includegraphics[width=0.5\textwidth]{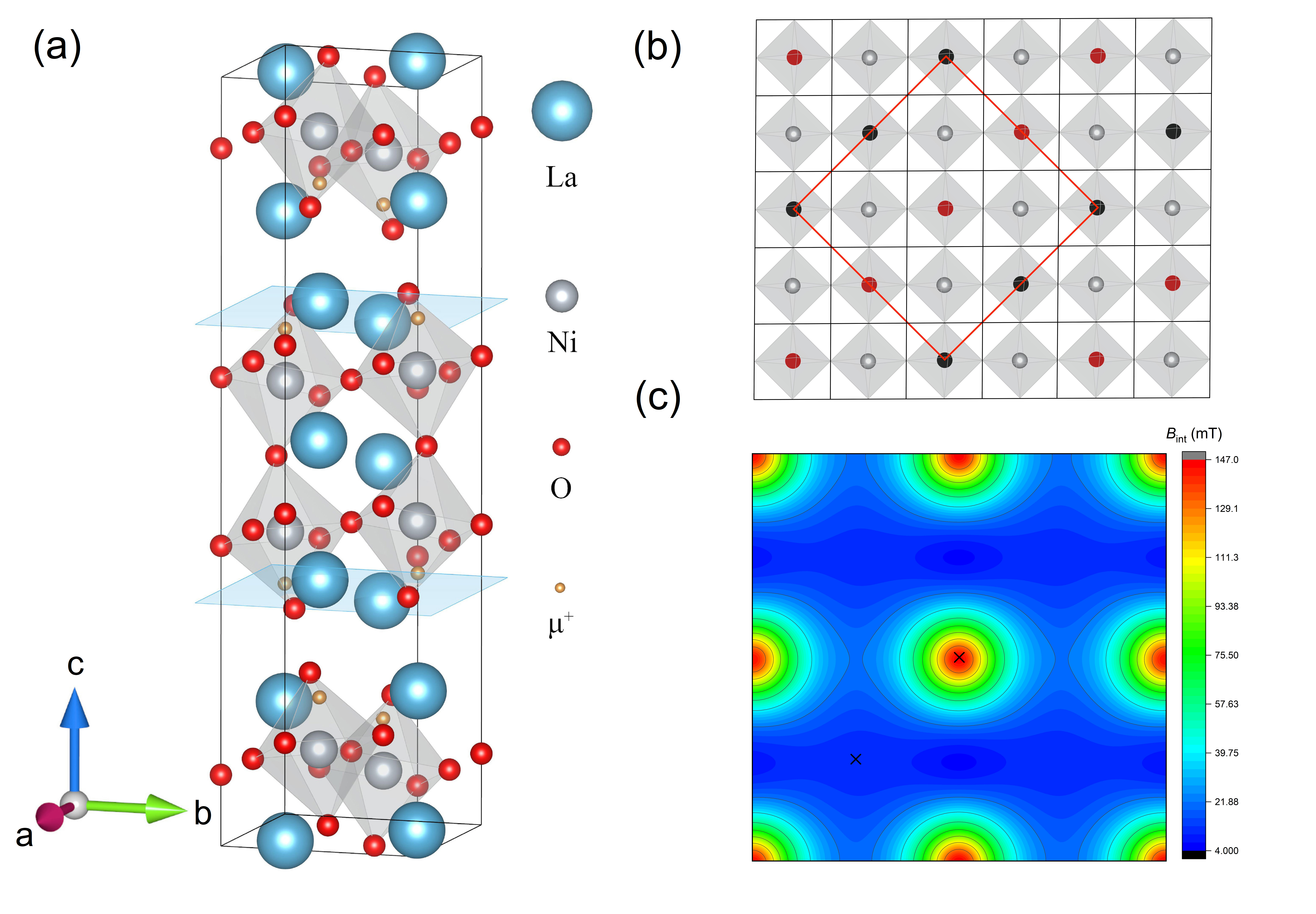}
			\caption{\label{fig:Fig4}(a) Candidate muon stopping sites calculated by DFT+$\mu$. The La(2) plane is highlighted. (b) Spin structure for the spin-charge strip proposed in Ref~\cite{RIXS}. The black, grey and red circles represent spin down $\rm{Ni^{2+}}$, spin up $\rm{Ni^{2+}}$ and spinless $\rm{Ni^{3+}}$, respectively. The spin polarization is along $c$-axis and antiferromagnetically coupled between neighboring NiO$_2$ layers. The magnetic unit cell is marked with red lines. (c) Dipolar field distribution in the muon residence plane ($a, b, 0.166$) using spin configuration shown in (b). The field distribution with $m_{\rm{Ni}}=0.22\ \rm{\mu_{B}}$ is shown in one magnetic unit cell. The $x/y$-axis is along [100]/[010] crystal orientation. Two crosses stand for two magnetically inequivalent sites in the magnetic unit cell with $B_1=144.6\ \rm{mT}$ and $B_2=11.5\ \rm{mT}$, respectively.}
		\end{center}
	\end{figure}

	\textit{Discussion.}---The candidate muon stopping sites are calculated with \texttt{MuFinder}~\cite{MuFinder} application using the \texttt{CASTEP} programme~\cite{Castep}. The internal magnetic field is calculated with \texttt{DipoleCal}~\cite{DipoleCal}. The details of calculation are described in supplementary materials~\cite{Supmat}. Candidate muon stopping site is shown in Fig.~\ref{fig:Fig4}(a). The source of the internal field is attributed to the ordered nickel moments in the vicinity of muon. The magnitude of nickel moments is estimated by the dipolar field strength at the candidate muon stopping sites with magnetic structures proposed in recent works~\cite{RIXS,NMR2024,NPD2024}. $m_{\rm{Ni}}\approx0.22/0.42\ \rm{\mu_{B}}$ when the magnetic moments align along $c/ab$-axis. The estimated moments are consistent with the value reported in Ref.~\cite{Zurab}, but does not seem to be small enough to account for the absence of magnetic order in neutron scattering experiment~\cite{NPD2024}. This may put the coherence length of magnetism in \LNOO\ on the `short' side of neutron scattering, since \musr\ is sensitive to short-range magnetism~\cite{Bonilla2011}. The in-plane rotation of nickel moments will not significantly change the field strength sensed by muons. A field distribution with both high and low field strengths can be reproduced with magnetic structure considering charge-stripes. Though the field distribution will be slightly influenced by the stacking pattern between bilayer NiO$_2$ planes along $c$-axis, this will not qualitatively change the field distribution in the plane where muon stops. The dipolar field strength at two magnetically inequivalent sites is qualitatively consistent with the value observed in the ZF-\musr\ experiment. In addition, the field strength near the muon stopping sites does not change rapidly, which is consistent with the slow decay of $B_{\rm{fast1}}$ and $B_{\rm{slow}}$. More spin structures are discussed in supplementary material~\cite{Supmat}. 
	
	Next, we will discuss the origin of fast precessing 2 component. $\lambda_{\rm{fast2}}$ is an order of magnitude larger than $\lambda_{\rm{fast1}}$ and $\lambda_{\rm{slow}}$~\cite{Supmat}. This indicates the internal field of $B_{\rm{fast2}}$ is rather inhomogenous, which cannot be attributed to magnetically inequivalent sites considering the cell symmetry of candidate sites. We therefore attribute $B_{\rm{fast2}}$ to the internal field inhomogeneity induced by oxygen vacancies~\cite{NMR2024}. It is reported that the inner apical oxygen vacancies will induce inhomogenous internal magnetic field at La(2) (highlighted in Fig.~\ref{fig:Fig4}(a)), which is close to the candidate muon stopping site. The inhomogenous internal field can also account for the broad magnetic transition width observed in wTF measurement and the absence of transition in bulk property measurements~\cite{Zhang1994,Taniguchi1995,Wu2001}. The oxygen deficiency of our sample is about 1\% as revealed by TGA analysis. Assuming that all vacancies are located at the inner oxygen site~\cite{MEP}, the volume fraction of oxygen vacancies is estimated to be approximately 10\%. However, this value seems to be too small to account for the relative ratio of $B_{\rm{fast2}}$, which accounts for $\sim$70\% of fast precessing signals at low temperature. It should be mentioned that the current calculations are based on the perfect crystal structure. The random oxygen vacancies in real system can modify the local crystal structure~\cite{MEP}, thus modify muon stopping sites and local field distribution.
	
	\textit{Conclusions.}---In conclusion, the present ZF-\musr\ measurement and wTF-\musr\ measurements confirm the bulk commensurate magnetism in polycrystalline \LNOO\ at ambient pressure. The internal magnetic field distribution is qualitatively consistent with spin configurations with charge stripes. In addition, a small quantity of oxygen deficiencies can greatly broaden the internal magnetic field distribution sensed by muons. The findings of this work can promote a comprehensive understanding of the superconducting mechanism of \LNO\ at high pressure.
	
	{\textit{Acknowledgments--} 
		We are grateful to G. D. Morris, B. Hitti, and D. Arsenau of the TRIUMF CMMS for assistance during the experiments. This research was funded by the National Key Research and Development Program of China, No.~2022YFA1402203, the National Natural Science Foundations of China, No.~12174065, and the Shanghai Municipal Science and Technology Major Project Grant, No.~2019SHZDZX01. Y. F. Guo acknowledges the National Key R\&D Program of China (Grant No. 2023YFA1406100) and the Double First-Class Initiative Fund of ShanghaiTech University.


\begin{thebibliography}{63}%
	\makeatletter
	\providecommand \@ifxundefined [1]{%
		\@ifx{#1\undefined}
	}%
	\providecommand \@ifnum [1]{%
		\ifnum #1\expandafter \@firstoftwo
		\else \expandafter \@secondoftwo
		\fi
	}%
	\providecommand \@ifx [1]{%
		\ifx #1\expandafter \@firstoftwo
		\else \expandafter \@secondoftwo
		\fi
	}%
	\providecommand \natexlab [1]{#1}%
	\providecommand \enquote  [1]{``#1''}%
	\providecommand \bibnamefont  [1]{#1}%
	\providecommand \bibfnamefont [1]{#1}%
	\providecommand \citenamefont [1]{#1}%
	\providecommand \href@noop [0]{\@secondoftwo}%
	\providecommand \href [0]{\begingroup \@sanitize@url \@href}%
	\providecommand \@href[1]{\@@startlink{#1}\@@href}%
	\providecommand \@@href[1]{\endgroup#1\@@endlink}%
	\providecommand \@sanitize@url [0]{\catcode `\\12\catcode `\$12\catcode
		`\&12\catcode `\#12\catcode `\^12\catcode `\_12\catcode `\%12\relax}%
	\providecommand \@@startlink[1]{}%
	\providecommand \@@endlink[0]{}%
	\providecommand \url  [0]{\begingroup\@sanitize@url \@url }%
	\providecommand \@url [1]{\endgroup\@href {#1}{\urlprefix }}%
	\providecommand \urlprefix  [0]{URL }%
	\providecommand \Eprint [0]{\href }%
	\providecommand \doibase [0]{https://doi.org/}%
	\providecommand \selectlanguage [0]{\@gobble}%
	\providecommand \bibinfo  [0]{\@secondoftwo}%
	\providecommand \bibfield  [0]{\@secondoftwo}%
	\providecommand \translation [1]{[#1]}%
	\providecommand \BibitemOpen [0]{}%
	\providecommand \bibitemStop [0]{}%
	\providecommand \bibitemNoStop [0]{.\EOS\space}%
	\providecommand \EOS [0]{\spacefactor3000\relax}%
	\providecommand \BibitemShut  [1]{\csname bibitem#1\endcsname}%
	\let\auto@bib@innerbib\@empty
	\bibitem [{\citenamefont {Sun}\ \emph {et~al.}(2023)\citenamefont {Sun},
		\citenamefont {Huo}, \citenamefont {Hu}, \citenamefont {Li}, \citenamefont
		{Liu}, \citenamefont {Han}, \citenamefont {Tang}, \citenamefont {Mao},
		\citenamefont {Yang}, \citenamefont {Wang}, \citenamefont {Cheng},
		\citenamefont {Yao}, \citenamefont {Zhang},\ and\ \citenamefont
		{Wang}}]{Sun2023}%
	\BibitemOpen
	\bibfield  {author} {\bibinfo {author} {\bibfnamefont {H.}~\bibnamefont
			{Sun}}, \bibinfo {author} {\bibfnamefont {M.}~\bibnamefont {Huo}}, \bibinfo
		{author} {\bibfnamefont {X.}~\bibnamefont {Hu}}, \bibinfo {author}
		{\bibfnamefont {J.}~\bibnamefont {Li}}, \bibinfo {author} {\bibfnamefont
			{Z.}~\bibnamefont {Liu}}, \bibinfo {author} {\bibfnamefont {Y.}~\bibnamefont
			{Han}}, \bibinfo {author} {\bibfnamefont {L.}~\bibnamefont {Tang}}, \bibinfo
		{author} {\bibfnamefont {Z.}~\bibnamefont {Mao}}, \bibinfo {author}
		{\bibfnamefont {P.}~\bibnamefont {Yang}}, \bibinfo {author} {\bibfnamefont
			{B.}~\bibnamefont {Wang}}, \bibinfo {author} {\bibfnamefont {J.}~\bibnamefont
			{Cheng}}, \bibinfo {author} {\bibfnamefont {D.-X.}\ \bibnamefont {Yao}},
		\bibinfo {author} {\bibfnamefont {G.-M.}\ \bibnamefont {Zhang}},\ and\
		\bibinfo {author} {\bibfnamefont {M.}~\bibnamefont {Wang}},\ }\bibfield
	{title} {\bibinfo {title} {{Signatures of superconductivity near 80 K in a
				nickelate under high pressure}},\ }\href
	{https://doi.org/10.1038/s41586-023-06408-7} {\bibfield  {journal} {\bibinfo
			{journal} {Nature(London)}\ }\textbf {\bibinfo {volume} {621}},\ \bibinfo
		{pages} {493} (\bibinfo {year} {2023})}\BibitemShut {NoStop}%
	\bibitem [{\citenamefont {Hou}\ \emph {et~al.}(2023)\citenamefont {Hou},
		\citenamefont {Yang}, \citenamefont {Liu}, \citenamefont {Li}, \citenamefont
		{Shan}, \citenamefont {Ma}, \citenamefont {Wang}, \citenamefont {Wang},
		\citenamefont {Guo}, \citenamefont {Sun}, \citenamefont {Uwatoko},
		\citenamefont {Wang}, \citenamefont {Zhang}, \citenamefont {Wang},\ and\
		\citenamefont {Cheng}}]{Hou2023}%
	\BibitemOpen
	\bibfield  {author} {\bibinfo {author} {\bibfnamefont {J.}~\bibnamefont
			{Hou}}, \bibinfo {author} {\bibfnamefont {P.-T.}\ \bibnamefont {Yang}},
		\bibinfo {author} {\bibfnamefont {Z.-Y.}\ \bibnamefont {Liu}}, \bibinfo
		{author} {\bibfnamefont {J.-Y.}\ \bibnamefont {Li}}, \bibinfo {author}
		{\bibfnamefont {P.-F.}\ \bibnamefont {Shan}}, \bibinfo {author}
		{\bibfnamefont {L.}~\bibnamefont {Ma}}, \bibinfo {author} {\bibfnamefont
			{G.}~\bibnamefont {Wang}}, \bibinfo {author} {\bibfnamefont {N.-N.}\
			\bibnamefont {Wang}}, \bibinfo {author} {\bibfnamefont {H.-Z.}\ \bibnamefont
			{Guo}}, \bibinfo {author} {\bibfnamefont {J.-P.}\ \bibnamefont {Sun}},
		\bibinfo {author} {\bibfnamefont {Y.}~\bibnamefont {Uwatoko}}, \bibinfo
		{author} {\bibfnamefont {M.}~\bibnamefont {Wang}}, \bibinfo {author}
		{\bibfnamefont {G.-M.}\ \bibnamefont {Zhang}}, \bibinfo {author}
		{\bibfnamefont {B.-S.}\ \bibnamefont {Wang}},\ and\ \bibinfo {author}
		{\bibfnamefont {J.-G.}\ \bibnamefont {Cheng}},\ }\bibfield  {title} {\bibinfo
		{title} {{Emergence of High-Temperature Superconducting Phase in Pressurized
				${\mathrm{La}}_{3}{\mathrm{Ni}}_{2}{\mathrm{O}}_{7}$ Crystals}},\ }\href
	{https://doi.org/10.1088/0256-307X/40/11/117302} {\bibfield  {journal}
		{\bibinfo  {journal} {Chin. Phys. Lett.}\ }\textbf {\bibinfo {volume} {40}},\
		\bibinfo {pages} {117302} (\bibinfo {year} {2023})}\BibitemShut {NoStop}%
	\bibitem [{\citenamefont {{Zhang, Mingxin and Pei, Cuiying and Wang, Qi and
				Zhao, Yi and Li, Changhua and Cao, Weizheng and Zhu, Shihao and Wu, Juefei
				and Qi, Yanpeng}}()}]{Zhang2023}%
	\BibitemOpen
	\bibfield  {author} {\bibinfo {author} {\bibnamefont {{Zhang, Mingxin and
					Pei, Cuiying and Wang, Qi and Zhao, Yi and Li, Changhua and Cao, Weizheng and
					Zhu, Shihao and Wu, Juefei and Qi, Yanpeng}}},\ }\bibfield  {title} {\bibinfo
		{title} {{Effects of Pressure and Doping on Ruddlesden-Popper phases
				$\mathrm{La}_{n+1}\mathrm{Ni}_n\mathrm{O}_{3n+1}$}},\ }\href@noop {} {\
	}\Eprint {https://arxiv.org/abs/2309.01651} {arXiv:2309.01651} \BibitemShut
	{NoStop}%
	\bibitem [{\citenamefont {Wang}\ \emph {et~al.}(2024)\citenamefont {Wang},
		\citenamefont {Wang}, \citenamefont {Shen}, \citenamefont {Hou},
		\citenamefont {Ma}, \citenamefont {Shi}, \citenamefont {Ren}, \citenamefont
		{Gu}, \citenamefont {Ma}, \citenamefont {Yang}, \citenamefont {Liu},
		\citenamefont {Guo}, \citenamefont {Sun}, \citenamefont {Zhang},
		\citenamefont {Calder}, \citenamefont {Yan}, \citenamefont {Wang},
		\citenamefont {Uwatoko},\ and\ \citenamefont {Cheng}}]{Wang2023}%
	\BibitemOpen
	\bibfield  {author} {\bibinfo {author} {\bibfnamefont {G.}~\bibnamefont
			{Wang}}, \bibinfo {author} {\bibfnamefont {N.~N.}\ \bibnamefont {Wang}},
		\bibinfo {author} {\bibfnamefont {X.~L.}\ \bibnamefont {Shen}}, \bibinfo
		{author} {\bibfnamefont {J.}~\bibnamefont {Hou}}, \bibinfo {author}
		{\bibfnamefont {L.}~\bibnamefont {Ma}}, \bibinfo {author} {\bibfnamefont
			{L.~F.}\ \bibnamefont {Shi}}, \bibinfo {author} {\bibfnamefont {Z.~A.}\
			\bibnamefont {Ren}}, \bibinfo {author} {\bibfnamefont {Y.~D.}\ \bibnamefont
			{Gu}}, \bibinfo {author} {\bibfnamefont {H.~M.}\ \bibnamefont {Ma}}, \bibinfo
		{author} {\bibfnamefont {P.~T.}\ \bibnamefont {Yang}}, \bibinfo {author}
		{\bibfnamefont {Z.~Y.}\ \bibnamefont {Liu}}, \bibinfo {author} {\bibfnamefont
			{H.~Z.}\ \bibnamefont {Guo}}, \bibinfo {author} {\bibfnamefont {J.~P.}\
			\bibnamefont {Sun}}, \bibinfo {author} {\bibfnamefont {G.~M.}\ \bibnamefont
			{Zhang}}, \bibinfo {author} {\bibfnamefont {S.}~\bibnamefont {Calder}},
		\bibinfo {author} {\bibfnamefont {J.-Q.}\ \bibnamefont {Yan}}, \bibinfo
		{author} {\bibfnamefont {B.~S.}\ \bibnamefont {Wang}}, \bibinfo {author}
		{\bibfnamefont {Y.}~\bibnamefont {Uwatoko}},\ and\ \bibinfo {author}
		{\bibfnamefont {J.-G.}\ \bibnamefont {Cheng}},\ }\bibfield  {title} {\bibinfo
		{title} {{Pressure-Induced Superconductivity In Polycrystalline
				${\mathrm{La}}_{3}{\mathrm{Ni}}_{2}{\mathrm{O}}_{7\ensuremath{-}\ensuremath{\delta}}$}},\
	}\href {https://doi.org/10.1103/PhysRevX.14.011040} {\bibfield  {journal}
		{\bibinfo  {journal} {Phys. Rev. X}\ }\textbf {\bibinfo {volume} {14}},\
		\bibinfo {pages} {011040} (\bibinfo {year} {2024})}\BibitemShut {NoStop}%
	\bibitem [{\citenamefont {Bednorz}\ and\ \citenamefont
		{Müller}(1986)}]{Bednorz1986}%
	\BibitemOpen
	\bibfield  {author} {\bibinfo {author} {\bibfnamefont {J.~G.}\ \bibnamefont
			{Bednorz}}\ and\ \bibinfo {author} {\bibfnamefont {K.~A.}\ \bibnamefont
			{Müller}},\ }\bibfield  {title} {\bibinfo {title} {{Possible high $T_c$
				superconductivity in the Ba-La-Cu-O system}},\ }\href
	{https://doi.org/10.1007/BF01303701} {\bibfield  {journal} {\bibinfo
			{journal} {Z. Naturforsch. B}\ }\textbf {\bibinfo {volume} {64}},\ \bibinfo
		{pages} {189} (\bibinfo {year} {1986})}\BibitemShut {NoStop}%
	\bibitem [{\citenamefont {Wu}\ \emph {et~al.}(1987)\citenamefont {Wu},
		\citenamefont {Ashburn}, \citenamefont {Torng}, \citenamefont {Hor},
		\citenamefont {Meng}, \citenamefont {Gao}, \citenamefont {Huang},
		\citenamefont {Wang},\ and\ \citenamefont {Chu}}]{Wu1987}%
	\BibitemOpen
	\bibfield  {author} {\bibinfo {author} {\bibfnamefont {M.~K.}\ \bibnamefont
			{Wu}}, \bibinfo {author} {\bibfnamefont {J.~R.}\ \bibnamefont {Ashburn}},
		\bibinfo {author} {\bibfnamefont {C.~J.}\ \bibnamefont {Torng}}, \bibinfo
		{author} {\bibfnamefont {P.~H.}\ \bibnamefont {Hor}}, \bibinfo {author}
		{\bibfnamefont {R.~L.}\ \bibnamefont {Meng}}, \bibinfo {author}
		{\bibfnamefont {L.}~\bibnamefont {Gao}}, \bibinfo {author} {\bibfnamefont
			{Z.~J.}\ \bibnamefont {Huang}}, \bibinfo {author} {\bibfnamefont {Y.~Q.}\
			\bibnamefont {Wang}},\ and\ \bibinfo {author} {\bibfnamefont {C.~W.}\
			\bibnamefont {Chu}},\ }\bibfield  {title} {\bibinfo {title}
		{{Superconductivity at 93 K in a new mixed-phase Y-Ba-Cu-O compound system at
				ambient pressure}},\ }\href {https://doi.org/10.1103/PhysRevLett.58.908}
	{\bibfield  {journal} {\bibinfo  {journal} {Phys. Rev. Lett.}\ }\textbf
		{\bibinfo {volume} {58}},\ \bibinfo {pages} {908} (\bibinfo {year}
		{1987})}\BibitemShut {NoStop}%
	\bibitem [{\citenamefont {Maeda}\ \emph {et~al.}(1988)\citenamefont {Maeda},
		\citenamefont {Tanaka}, \citenamefont {Fukutomi},\ and\ \citenamefont
		{Asano}}]{Maeda1988}%
	\BibitemOpen
	\bibfield  {author} {\bibinfo {author} {\bibfnamefont {H.}~\bibnamefont
			{Maeda}}, \bibinfo {author} {\bibfnamefont {Y.}~\bibnamefont {Tanaka}},
		\bibinfo {author} {\bibfnamefont {M.}~\bibnamefont {Fukutomi}},\ and\
		\bibinfo {author} {\bibfnamefont {T.}~\bibnamefont {Asano}},\ }\bibfield
	{title} {\bibinfo {title} {{A New High-Tc Oxide Superconductor without a Rare
				Earth Element}},\ }\href {https://doi.org/10.1143/JJAP.27.L209} {\bibfield
		{journal} {\bibinfo  {journal} {Jpn. J. Appl. Phys.}\ }\textbf {\bibinfo
			{volume} {27}},\ \bibinfo {pages} {L209} (\bibinfo {year}
		{1988})}\BibitemShut {NoStop}%
	\bibitem [{\citenamefont {Sheng}\ and\ \citenamefont
		{Hermann}(1988)}]{Sheng1988}%
	\BibitemOpen
	\bibfield  {author} {\bibinfo {author} {\bibfnamefont {Z.~Z.}\ \bibnamefont
			{Sheng}}\ and\ \bibinfo {author} {\bibfnamefont {A.~M.}\ \bibnamefont
			{Hermann}},\ }\bibfield  {title} {\bibinfo {title} {{Bulk superconductivity
				at 120 K in the Tl-Ca/Ba-Cu-O system}},\ }\href
	{https://doi.org/10.1038/332138a0} {\bibfield  {journal} {\bibinfo  {journal}
			{Nature(London)}\ }\textbf {\bibinfo {volume} {332}},\ \bibinfo {pages} {138}
		(\bibinfo {year} {1988})}\BibitemShut {NoStop}%
	\bibitem [{\citenamefont {Ge}\ \emph {et~al.}(2015)\citenamefont {Ge},
		\citenamefont {Liu}, \citenamefont {Liu}, \citenamefont {Gao}, \citenamefont
		{Qian}, \citenamefont {Xue}, \citenamefont {Liu},\ and\ \citenamefont
		{Jia}}]{Ge2015}%
	\BibitemOpen
	\bibfield  {author} {\bibinfo {author} {\bibfnamefont {J.-F.}\ \bibnamefont
			{Ge}}, \bibinfo {author} {\bibfnamefont {Z.-L.}\ \bibnamefont {Liu}},
		\bibinfo {author} {\bibfnamefont {C.}~\bibnamefont {Liu}}, \bibinfo {author}
		{\bibfnamefont {C.-L.}\ \bibnamefont {Gao}}, \bibinfo {author} {\bibfnamefont
			{D.}~\bibnamefont {Qian}}, \bibinfo {author} {\bibfnamefont {Q.-K.}\
			\bibnamefont {Xue}}, \bibinfo {author} {\bibfnamefont {Y.}~\bibnamefont
			{Liu}},\ and\ \bibinfo {author} {\bibfnamefont {J.-F.}\ \bibnamefont {Jia}},\
	}\bibfield  {title} {\bibinfo {title} {{Superconductivity above 100 K in
				single-layer FeSe films on doped $\mathrm{SrTiO}_{3}$}},\ }\href
	{https://doi.org/10.1038/nmat4153} {\bibfield  {journal} {\bibinfo  {journal}
			{Nat. Mater.}\ }\textbf {\bibinfo {volume} {14}},\ \bibinfo {pages} {285}
		(\bibinfo {year} {2015})}\BibitemShut {NoStop}%
	\bibitem [{\citenamefont {Eremets}\ \emph {et~al.}(2022)\citenamefont
		{Eremets}, \citenamefont {Minkov}, \citenamefont {Drozdov}, \citenamefont
		{Kong}, \citenamefont {Ksenofontov}, \citenamefont {Shylin}, \citenamefont
		{Bud’ko}, \citenamefont {Prozorov}, \citenamefont {Balakirev},
		\citenamefont {Sun}, \citenamefont {Mozaffari},\ and\ \citenamefont
		{Balicas}}]{Eremets2022}%
	\BibitemOpen
	\bibfield  {author} {\bibinfo {author} {\bibfnamefont {M.~I.}\ \bibnamefont
			{Eremets}}, \bibinfo {author} {\bibfnamefont {V.~S.}\ \bibnamefont {Minkov}},
		\bibinfo {author} {\bibfnamefont {A.~P.}\ \bibnamefont {Drozdov}}, \bibinfo
		{author} {\bibfnamefont {P.~P.}\ \bibnamefont {Kong}}, \bibinfo {author}
		{\bibfnamefont {V.}~\bibnamefont {Ksenofontov}}, \bibinfo {author}
		{\bibfnamefont {S.~I.}\ \bibnamefont {Shylin}}, \bibinfo {author}
		{\bibfnamefont {S.~L.}\ \bibnamefont {Bud’ko}}, \bibinfo {author}
		{\bibfnamefont {R.}~\bibnamefont {Prozorov}}, \bibinfo {author}
		{\bibfnamefont {F.~F.}\ \bibnamefont {Balakirev}}, \bibinfo {author}
		{\bibfnamefont {D.}~\bibnamefont {Sun}}, \bibinfo {author} {\bibfnamefont
			{S.}~\bibnamefont {Mozaffari}},\ and\ \bibinfo {author} {\bibfnamefont
			{L.}~\bibnamefont {Balicas}},\ }\bibfield  {title} {\bibinfo {title}
		{{High-Temperature Superconductivity in Hydrides: Experimental Evidence and
				Details}},\ }\href {https://doi.org/10.1007/s10948-022-06148-1} {\bibfield
		{journal} {\bibinfo  {journal} {J. Supercond. Novel Magn.}\ }\textbf
		{\bibinfo {volume} {35}},\ \bibinfo {pages} {965} (\bibinfo {year}
		{2022})}\BibitemShut {NoStop}%
	\bibitem [{\citenamefont {Li}\ \emph {et~al.}()\citenamefont {Li},
		\citenamefont {Ma}, \citenamefont {Zhang}, \citenamefont {Huang},
		\citenamefont {Huang}, \citenamefont {Huo}, \citenamefont {Hu}, \citenamefont
		{Dong}, \citenamefont {He}, \citenamefont {Liao}, \citenamefont {Chen},
		\citenamefont {Xie}, \citenamefont {Sun},\ and\ \citenamefont
		{Wang}}]{Jing2024}%
	\BibitemOpen
	\bibfield  {author} {\bibinfo {author} {\bibfnamefont {J.}~\bibnamefont
			{Li}}, \bibinfo {author} {\bibfnamefont {P.}~\bibnamefont {Ma}}, \bibinfo
		{author} {\bibfnamefont {H.}~\bibnamefont {Zhang}}, \bibinfo {author}
		{\bibfnamefont {X.}~\bibnamefont {Huang}}, \bibinfo {author} {\bibfnamefont
			{C.}~\bibnamefont {Huang}}, \bibinfo {author} {\bibfnamefont
			{M.}~\bibnamefont {Huo}}, \bibinfo {author} {\bibfnamefont {D.}~\bibnamefont
			{Hu}}, \bibinfo {author} {\bibfnamefont {Z.}~\bibnamefont {Dong}}, \bibinfo
		{author} {\bibfnamefont {C.}~\bibnamefont {He}}, \bibinfo {author}
		{\bibfnamefont {J.}~\bibnamefont {Liao}}, \bibinfo {author} {\bibfnamefont
			{X.}~\bibnamefont {Chen}}, \bibinfo {author} {\bibfnamefont {T.}~\bibnamefont
			{Xie}}, \bibinfo {author} {\bibfnamefont {H.}~\bibnamefont {Sun}},\ and\
		\bibinfo {author} {\bibfnamefont {M.}~\bibnamefont {Wang}},\ }\bibfield
	{title} {\bibinfo {title} {{Pressure-driven right-triangle shape
				superconductivity in bilayer nickelate
				$\mathrm{La}_3\mathrm{Ni}_2\mathrm{O}_{7}$}},\ }\href@noop {} {\ }\Eprint
	{https://arxiv.org/abs/2404.11369} {arXiv:2404.11369} \BibitemShut {NoStop}%
	\bibitem [{\citenamefont {{Luo, Zhihui and Hu, Xunwu and Wang, Meng and Wú,
				Wéi and Yao, Dao-Xin}}(2023)}]{Luo2023}%
	\BibitemOpen
	\bibfield  {author} {\bibinfo {author} {\bibnamefont {{Luo, Zhihui and Hu,
					Xunwu and Wang, Meng and Wú, Wéi and Yao, Dao-Xin}}},\ }\bibfield  {title}
	{\bibinfo {title} {{Bilayer Two-Orbital Model of
				$\mathrm{L}{\mathrm{a}}_{3}\mathrm{N}{\mathrm{i}}_{2}{\mathrm{O}}_{7}$ under
				Pressure}},\ }\href {https://doi.org/10.1103/PhysRevLett.131.126001}
	{\bibfield  {journal} {\bibinfo  {journal} {Phys. Rev. Lett.}\ }\textbf
		{\bibinfo {volume} {131}},\ \bibinfo {pages} {126001} (\bibinfo {year}
		{2023})}\BibitemShut {NoStop}%
	\bibitem [{\citenamefont {Qin}\ and\ \citenamefont {Yang}(2023)}]{Qin2023}%
	\BibitemOpen
	\bibfield  {author} {\bibinfo {author} {\bibfnamefont {Q.}~\bibnamefont
			{Qin}}\ and\ \bibinfo {author} {\bibfnamefont {Y.-f.}\ \bibnamefont {Yang}},\
	}\bibfield  {title} {\bibinfo {title} {{High-${T}_{c}$ superconductivity by
				mobilizing local spin singlets and possible route to higher ${T}_{c}$ in
				pressurized ${\mathrm{La}}_{3}{\mathrm{Ni}}_{2}{\mathrm{O}}_{7}$}},\ }\href
	{https://doi.org/10.1103/PhysRevB.108.L140504} {\bibfield  {journal}
		{\bibinfo  {journal} {Phys. Rev. B}\ }\textbf {\bibinfo {volume} {108}},\
		\bibinfo {pages} {L140504} (\bibinfo {year} {2023})}\BibitemShut {NoStop}%
	\bibitem [{\citenamefont {Shilenko}\ and\ \citenamefont
		{Leonov}(2023)}]{Shilenko2023}%
	\BibitemOpen
	\bibfield  {author} {\bibinfo {author} {\bibfnamefont {D.~A.}\ \bibnamefont
			{Shilenko}}\ and\ \bibinfo {author} {\bibfnamefont {I.~V.}\ \bibnamefont
			{Leonov}},\ }\bibfield  {title} {\bibinfo {title} {{Correlated electronic
				structure, orbital-selective behavior, and magnetic correlations in
				double-layer ${\mathrm{La}}_{3}{\mathrm{Ni}}_{2}{\mathrm{O}}_{7}$ under
				pressure}},\ }\href {https://doi.org/10.1103/PhysRevB.108.125105} {\bibfield
		{journal} {\bibinfo  {journal} {Phys. Rev. B}\ }\textbf {\bibinfo {volume}
			{108}},\ \bibinfo {pages} {125105} (\bibinfo {year} {2023})}\BibitemShut
	{NoStop}%
	\bibitem [{\citenamefont {Yang}\ \emph
		{et~al.}(2023{\natexlab{a}})\citenamefont {Yang}, \citenamefont {Wang},\ and\
		\citenamefont {Wang}}]{Yang2023_theory}%
	\BibitemOpen
	\bibfield  {author} {\bibinfo {author} {\bibfnamefont {Q.-G.}\ \bibnamefont
			{Yang}}, \bibinfo {author} {\bibfnamefont {D.}~\bibnamefont {Wang}},\ and\
		\bibinfo {author} {\bibfnamefont {Q.-H.}\ \bibnamefont {Wang}},\ }\bibfield
	{title} {\bibinfo {title} {{Possible
				${s}_{\ifmmode\pm\else\textpm\fi{}}$-wave superconductivity in
				${\mathrm{La}}_{3}{\mathrm{Ni}}_{2}{\mathrm{O}}_{7}$}},\ }\href
	{https://doi.org/10.1103/PhysRevB.108.L140505} {\bibfield  {journal}
		{\bibinfo  {journal} {Phys. Rev. B}\ }\textbf {\bibinfo {volume} {108}},\
		\bibinfo {pages} {L140505} (\bibinfo {year}
		{2023}{\natexlab{a}})}\BibitemShut {NoStop}%
	\bibitem [{\citenamefont {Yang}\ \emph
		{et~al.}(2023{\natexlab{b}})\citenamefont {Yang}, \citenamefont {Zhang},\
		and\ \citenamefont {Zhang}}]{YYF2023}%
	\BibitemOpen
	\bibfield  {author} {\bibinfo {author} {\bibfnamefont {Y.-f.}\ \bibnamefont
			{Yang}}, \bibinfo {author} {\bibfnamefont {G.-M.}\ \bibnamefont {Zhang}},\
		and\ \bibinfo {author} {\bibfnamefont {F.-C.}\ \bibnamefont {Zhang}},\
	}\bibfield  {title} {\bibinfo {title} {{Interlayer valence bonds and
				two-component theory for high-${T}_{c}$ superconductivity of
				${\mathrm{La}}_{3}{\mathrm{Ni}}_{2}{\mathrm{O}}_{7}$ under pressure}},\
	}\href {https://doi.org/10.1103/PhysRevB.108.L201108} {\bibfield  {journal}
		{\bibinfo  {journal} {Phys. Rev. B}\ }\textbf {\bibinfo {volume} {108}},\
		\bibinfo {pages} {L201108} (\bibinfo {year}
		{2023}{\natexlab{b}})}\BibitemShut {NoStop}%
	\bibitem [{\citenamefont {{Cao}}\ and\ \citenamefont {{Yang}}()}]{Cao2023}%
	\BibitemOpen
	\bibfield  {author} {\bibinfo {author} {\bibfnamefont {Y.}~\bibnamefont
			{{Cao}}}\ and\ \bibinfo {author} {\bibfnamefont {Y.-f.}\ \bibnamefont
			{{Yang}}},\ }\bibfield  {title} {\bibinfo {title} {{Flat bands promoted by
				Hund's rule coupling in the candidate double-layer high-temperature
				superconductor $\mathrm{La}_3\mathrm{Ni}_2\mathrm{O}_7$}},\ }\href@noop {} {\
	}\Eprint {https://arxiv.org/abs/2307.06806} {arXiv:2307.06806} \BibitemShut
	{NoStop}%
	\bibitem [{\citenamefont {Chen}\ \emph {et~al.}({\natexlab{a}})\citenamefont
		{Chen}, \citenamefont {Jiang}, \citenamefont {Li}, \citenamefont {Zhong},\
		and\ \citenamefont {Lu}}]{Chen2023}%
	\BibitemOpen
	\bibfield  {author} {\bibinfo {author} {\bibfnamefont {X.}~\bibnamefont
			{Chen}}, \bibinfo {author} {\bibfnamefont {P.}~\bibnamefont {Jiang}},
		\bibinfo {author} {\bibfnamefont {J.}~\bibnamefont {Li}}, \bibinfo {author}
		{\bibfnamefont {Z.}~\bibnamefont {Zhong}},\ and\ \bibinfo {author}
		{\bibfnamefont {Y.}~\bibnamefont {Lu}},\ }\bibfield  {title} {\bibinfo
		{title} {{Critical charge and spin instabilities in superconducting
				$\mathrm{La}_3\mathrm{Ni}_2\mathrm{O}_7$}},\ }\href@noop {} {\
		({\natexlab{a}})},\ \Eprint {https://arxiv.org/abs/2307.07154}
	{arXiv:2307.07154} \BibitemShut {NoStop}%
	\bibitem [{\citenamefont {Christiansson}\ \emph {et~al.}(2023)\citenamefont
		{Christiansson}, \citenamefont {Petocchi},\ and\ \citenamefont
		{Werner}}]{Christiansson2023}%
	\BibitemOpen
	\bibfield  {author} {\bibinfo {author} {\bibfnamefont {V.}~\bibnamefont
			{Christiansson}}, \bibinfo {author} {\bibfnamefont {F.}~\bibnamefont
			{Petocchi}},\ and\ \bibinfo {author} {\bibfnamefont {P.}~\bibnamefont
			{Werner}},\ }\bibfield  {title} {\bibinfo {title} {{Correlated Electronic
				Structure of ${\mathrm{La}}_{3}{\text{Ni}}_{2}{\mathrm{O}}_{7}$ under
				Pressure}},\ }\href {https://doi.org/10.1103/PhysRevLett.131.206501}
	{\bibfield  {journal} {\bibinfo  {journal} {Phys. Rev. Lett.}\ }\textbf
		{\bibinfo {volume} {131}},\ \bibinfo {pages} {206501} (\bibinfo {year}
		{2023})}\BibitemShut {NoStop}%
	\bibitem [{\citenamefont {Jiang}\ \emph {et~al.}()\citenamefont {Jiang},
		\citenamefont {Wang},\ and\ \citenamefont {Zhang}}]{Jiang2023}%
	\BibitemOpen
	\bibfield  {author} {\bibinfo {author} {\bibfnamefont {K.}~\bibnamefont
			{Jiang}}, \bibinfo {author} {\bibfnamefont {Z.}~\bibnamefont {Wang}},\ and\
		\bibinfo {author} {\bibfnamefont {F.-C.}\ \bibnamefont {Zhang}},\ }\bibfield
	{title} {\bibinfo {title} {{High Temperature Superconductivity in
				$\mathrm{La}_3\mathrm{Ni}_2\mathrm{O}_7$}},\ }\href@noop {} {\ }\Eprint
	{https://arxiv.org/abs/2308.06771} {arXiv:2308.06771} \BibitemShut {NoStop}%
	\bibitem [{\citenamefont {LaBollita}\ \emph {et~al.}()\citenamefont
		{LaBollita}, \citenamefont {Pardo}, \citenamefont {Norman},\ and\
		\citenamefont {Botana}}]{LaBollita2023}%
	\BibitemOpen
	\bibfield  {author} {\bibinfo {author} {\bibfnamefont {H.}~\bibnamefont
			{LaBollita}}, \bibinfo {author} {\bibfnamefont {V.}~\bibnamefont {Pardo}},
		\bibinfo {author} {\bibfnamefont {M.~R.}\ \bibnamefont {Norman}},\ and\
		\bibinfo {author} {\bibfnamefont {A.~S.}\ \bibnamefont {Botana}},\ }\bibfield
	{title} {\bibinfo {title} {{Electronic structure and magnetic properties of
				$\mathrm{La}_3\mathrm{Ni}_2\mathrm{O}_7$ under pressure}},\ }\href@noop {} {\
	}\Eprint {https://arxiv.org/abs/2309.17279} {arXiv:2309.17279} \BibitemShut
	{NoStop}%
	\bibitem [{\citenamefont {Shen}\ \emph {et~al.}(2023)\citenamefont {Shen},
		\citenamefont {Qin},\ and\ \citenamefont {Zhang}}]{Shen2023}%
	\BibitemOpen
	\bibfield  {author} {\bibinfo {author} {\bibfnamefont {Y.}~\bibnamefont
			{Shen}}, \bibinfo {author} {\bibfnamefont {M.}~\bibnamefont {Qin}},\ and\
		\bibinfo {author} {\bibfnamefont {G.-M.}\ \bibnamefont {Zhang}},\ }\bibfield
	{title} {\bibinfo {title} {{Effective Bi-Layer Model Hamiltonian and
				Density-Matrix Renormalization Group Study for the High-$T_c$
				Superconductivity in $\mathrm{La}_{3}\mathrm{Ni}_{2}\mathrm{O}_{7}$ under
				High Pressure}},\ }\href {https://doi.org/10.1088/0256-307X/40/12/127401}
	{\bibfield  {journal} {\bibinfo  {journal} {Chin. Phys. Lett.}\ }\textbf
		{\bibinfo {volume} {40}},\ \bibinfo {pages} {127401} (\bibinfo {year}
		{2023})}\BibitemShut {NoStop}%
	\bibitem [{\citenamefont {Zhang}\ \emph {et~al.}(2023)\citenamefont {Zhang},
		\citenamefont {Lin}, \citenamefont {Moreo}, \citenamefont {Maier},\ and\
		\citenamefont {Dagotto}}]{Zhang2023_theory}%
	\BibitemOpen
	\bibfield  {author} {\bibinfo {author} {\bibfnamefont {Y.}~\bibnamefont
			{Zhang}}, \bibinfo {author} {\bibfnamefont {L.-F.}\ \bibnamefont {Lin}},
		\bibinfo {author} {\bibfnamefont {A.}~\bibnamefont {Moreo}}, \bibinfo
		{author} {\bibfnamefont {T.~A.}\ \bibnamefont {Maier}},\ and\ \bibinfo
		{author} {\bibfnamefont {E.}~\bibnamefont {Dagotto}},\ }\bibfield  {title}
	{\bibinfo {title} {{Trends in electronic structures and
				${s}_{\ifmmode\pm\else\textpm\fi{}}$-wave pairing for the rare-earth series
				in bilayer nickelate superconductor
				${R}_{3}{\mathrm{Ni}}_{2}{\mathrm{O}}_{7}$}},\ }\href
	{https://doi.org/10.1103/PhysRevB.108.165141} {\bibfield  {journal} {\bibinfo
			{journal} {Phys. Rev. B}\ }\textbf {\bibinfo {volume} {108}},\ \bibinfo
		{pages} {165141} (\bibinfo {year} {2023})}\BibitemShut {NoStop}%
	\bibitem [{\citenamefont {Yi-Heng~Tian}()}]{Tian2023}%
	\BibitemOpen
	\bibfield  {author} {\bibinfo {author} {\bibfnamefont {J.-M. W. R.-Q. H.
				Z.-Y.~L.}\ \bibnamefont {Yi-Heng~Tian}, \bibfnamefont {Yin~Chen}},\
	}\bibfield  {title} {\bibinfo {title} {{Correlation Effects and Concomitant
				Two-Orbital $s\pm$-Wave Superconductivity in
				$\mathrm{La}_{3}\mathrm{Ni}_{2}\mathrm{O}_{7}$ under High Pressure}},\
	}\href@noop {} {\ }\Eprint {https://arxiv.org/abs/2308.09698}
	{arXiv:2308.09698} \BibitemShut {NoStop}%
	\bibitem [{\citenamefont {Sakakibara}\ \emph {et~al.}()\citenamefont
		{Sakakibara}, \citenamefont {Kitamine}, \citenamefont {Ochi},\ and\
		\citenamefont {Kuroki}}]{Sakakibara2023}%
	\BibitemOpen
	\bibfield  {author} {\bibinfo {author} {\bibfnamefont {H.}~\bibnamefont
			{Sakakibara}}, \bibinfo {author} {\bibfnamefont {N.}~\bibnamefont
			{Kitamine}}, \bibinfo {author} {\bibfnamefont {M.}~\bibnamefont {Ochi}},\
		and\ \bibinfo {author} {\bibfnamefont {K.}~\bibnamefont {Kuroki}},\
	}\bibfield  {title} {\bibinfo {title} {{Possible high $T_c$ superconductivity
				in $\mathrm{La}_3\mathrm{Ni}_2\mathrm{O}_7$ under high pressure through
				manifestation of a nearly-half-filled bilayer Hubbard model}},\ }\href@noop
	{} {\ }\Eprint {https://arxiv.org/abs/2306.06039} {arXiv:2306.06039}
	\BibitemShut {NoStop}%
	\bibitem [{\citenamefont {Gu}\ \emph {et~al.}()\citenamefont {Gu},
		\citenamefont {Le}, \citenamefont {Yang}, \citenamefont {Wu},\ and\
		\citenamefont {Hu}}]{Gu2023}%
	\BibitemOpen
	\bibfield  {author} {\bibinfo {author} {\bibfnamefont {Y.}~\bibnamefont
			{Gu}}, \bibinfo {author} {\bibfnamefont {C.}~\bibnamefont {Le}}, \bibinfo
		{author} {\bibfnamefont {Z.}~\bibnamefont {Yang}}, \bibinfo {author}
		{\bibfnamefont {X.}~\bibnamefont {Wu}},\ and\ \bibinfo {author}
		{\bibfnamefont {J.}~\bibnamefont {Hu}},\ }\bibfield  {title} {\bibinfo
		{title} {{Effective model and pairing tendency in bilayer Ni-based
				superconductor $\mathrm{La}_3\mathrm{Ni}_2\mathrm{O}_7$}},\ }\href@noop {} {\
	}\Eprint {https://arxiv.org/abs/2306.07275} {arXiv:2306.07275} \BibitemShut
	{NoStop}%
	\bibitem [{\citenamefont {Zhang}\ \emph {et~al.}()\citenamefont {Zhang},
		\citenamefont {Lin}, \citenamefont {Moreo}, \citenamefont {Maier},\ and\
		\citenamefont {Dagotto}}]{Zhang2023structural}%
	\BibitemOpen
	\bibfield  {author} {\bibinfo {author} {\bibfnamefont {Y.}~\bibnamefont
			{Zhang}}, \bibinfo {author} {\bibfnamefont {L.-F.}\ \bibnamefont {Lin}},
		\bibinfo {author} {\bibfnamefont {A.}~\bibnamefont {Moreo}}, \bibinfo
		{author} {\bibfnamefont {T.~A.}\ \bibnamefont {Maier}},\ and\ \bibinfo
		{author} {\bibfnamefont {E.}~\bibnamefont {Dagotto}},\ }\bibfield  {title}
	{\bibinfo {title} {{Structural phase transition, $s_{\pm}$-wave pairing and
				magnetic stripe order in the bilayered nickelate superconductor
				$\mathrm{La}_3\mathrm{Ni}_2\mathrm{O}_7$ under pressure}},\ }\href@noop {} {\
	}\Eprint {https://arxiv.org/abs/2307.15276} {arXiv:2307.15276} \BibitemShut
	{NoStop}%
	\bibitem [{\citenamefont {Yang}\ \emph {et~al.}()\citenamefont {Yang},
		\citenamefont {Sun}, \citenamefont {Hu}, \citenamefont {Xie}, \citenamefont
		{Miao}, \citenamefont {Luo}, \citenamefont {Chen}, \citenamefont {Liang},
		\citenamefont {Zhu}, \citenamefont {Qu}, \citenamefont {Chen}, \citenamefont
		{Huo}, \citenamefont {Huang}, \citenamefont {Zhang}, \citenamefont {Zhang},
		\citenamefont {Yang}, \citenamefont {Wang}, \citenamefont {Peng},
		\citenamefont {Mao}, \citenamefont {Liu}, \citenamefont {Xu}, \citenamefont
		{Qian}, \citenamefont {Yao}, \citenamefont {Wang}, \citenamefont {Zhao},\
		and\ \citenamefont {Zhou}}]{Yang2023}%
	\BibitemOpen
	\bibfield  {author} {\bibinfo {author} {\bibfnamefont {J.}~\bibnamefont
			{Yang}}, \bibinfo {author} {\bibfnamefont {H.}~\bibnamefont {Sun}}, \bibinfo
		{author} {\bibfnamefont {X.}~\bibnamefont {Hu}}, \bibinfo {author}
		{\bibfnamefont {Y.}~\bibnamefont {Xie}}, \bibinfo {author} {\bibfnamefont
			{T.}~\bibnamefont {Miao}}, \bibinfo {author} {\bibfnamefont {H.}~\bibnamefont
			{Luo}}, \bibinfo {author} {\bibfnamefont {H.}~\bibnamefont {Chen}}, \bibinfo
		{author} {\bibfnamefont {B.}~\bibnamefont {Liang}}, \bibinfo {author}
		{\bibfnamefont {W.}~\bibnamefont {Zhu}}, \bibinfo {author} {\bibfnamefont
			{G.}~\bibnamefont {Qu}}, \bibinfo {author} {\bibfnamefont {C.-Q.}\
			\bibnamefont {Chen}}, \bibinfo {author} {\bibfnamefont {M.}~\bibnamefont
			{Huo}}, \bibinfo {author} {\bibfnamefont {Y.}~\bibnamefont {Huang}}, \bibinfo
		{author} {\bibfnamefont {S.}~\bibnamefont {Zhang}}, \bibinfo {author}
		{\bibfnamefont {F.}~\bibnamefont {Zhang}}, \bibinfo {author} {\bibfnamefont
			{F.}~\bibnamefont {Yang}}, \bibinfo {author} {\bibfnamefont {Z.}~\bibnamefont
			{Wang}}, \bibinfo {author} {\bibfnamefont {Q.}~\bibnamefont {Peng}}, \bibinfo
		{author} {\bibfnamefont {H.}~\bibnamefont {Mao}}, \bibinfo {author}
		{\bibfnamefont {G.}~\bibnamefont {Liu}}, \bibinfo {author} {\bibfnamefont
			{Z.}~\bibnamefont {Xu}}, \bibinfo {author} {\bibfnamefont {T.}~\bibnamefont
			{Qian}}, \bibinfo {author} {\bibfnamefont {D.-X.}\ \bibnamefont {Yao}},
		\bibinfo {author} {\bibfnamefont {M.}~\bibnamefont {Wang}}, \bibinfo {author}
		{\bibfnamefont {L.}~\bibnamefont {Zhao}},\ and\ \bibinfo {author}
		{\bibfnamefont {X.~J.}\ \bibnamefont {Zhou}},\ }\bibfield  {title} {\bibinfo
		{title} {{Orbital-Dependent Electron Correlation in Double-Layer Nickelate
				$\mathrm{L}{\mathrm{a}}_{3}\mathrm{N}{\mathrm{i}}_{2}{\mathrm{O}}_{7}$}},\
	}\href@noop {} {\ }\Eprint {https://arxiv.org/abs/2309.01148}
	{arXiv:2309.01148} \BibitemShut {NoStop}%
	\bibitem [{\citenamefont {{Zhou}}\ \emph {et~al.}()\citenamefont {{Zhou}},
		\citenamefont {{Guo}}, \citenamefont {{Cai}}, \citenamefont {{Sun}},
		\citenamefont {{Wang}}, \citenamefont {{Zhao}}, \citenamefont {{Han}},
		\citenamefont {{Chen}}, \citenamefont {{Wu}}, \citenamefont {{Ding}},
		\citenamefont {{Wang}}, \citenamefont {{Xiang}}, \citenamefont {{Mao}},\ and\
		\citenamefont {{Sun}}}]{Zhou2023}%
	\BibitemOpen
	\bibfield  {author} {\bibinfo {author} {\bibfnamefont {Y.}~\bibnamefont
			{{Zhou}}}, \bibinfo {author} {\bibfnamefont {J.}~\bibnamefont {{Guo}}},
		\bibinfo {author} {\bibfnamefont {S.}~\bibnamefont {{Cai}}}, \bibinfo
		{author} {\bibfnamefont {H.}~\bibnamefont {{Sun}}}, \bibinfo {author}
		{\bibfnamefont {P.}~\bibnamefont {{Wang}}}, \bibinfo {author} {\bibfnamefont
			{J.}~\bibnamefont {{Zhao}}}, \bibinfo {author} {\bibfnamefont
			{J.}~\bibnamefont {{Han}}}, \bibinfo {author} {\bibfnamefont
			{X.}~\bibnamefont {{Chen}}}, \bibinfo {author} {\bibfnamefont
			{Q.}~\bibnamefont {{Wu}}}, \bibinfo {author} {\bibfnamefont {Y.}~\bibnamefont
			{{Ding}}}, \bibinfo {author} {\bibfnamefont {M.}~\bibnamefont {{Wang}}},
		\bibinfo {author} {\bibfnamefont {T.}~\bibnamefont {{Xiang}}}, \bibinfo
		{author} {\bibfnamefont {H.-k.}\ \bibnamefont {{Mao}}},\ and\ \bibinfo
		{author} {\bibfnamefont {L.}~\bibnamefont {{Sun}}},\ }\bibfield  {title}
	{\bibinfo {title} {{Evidence of filamentary superconductivity in pressurized
				$\mathrm{L}{\mathrm{a}}_{3}\mathrm{N}{\mathrm{i}}_{2}{\mathrm{O}}_{7}$ single
				crystals}},\ }\href@noop {} {\ }\Eprint {https://arxiv.org/abs/2311.12361}
	{arXiv:2311.12361} \BibitemShut {NoStop}%
	\bibitem [{\citenamefont {{Zhang, Yanan and Su, Dajun and Huang, Yanen and Sun,
				Hualei and Huo, Mengwu and Shan, Zhaoyang and Ye, Kaixin and Yang, Zihan and
				Li, Rui and Smidman, Michael and Wang, Meng and Jiao, Lin and Yuan,
				Huiqiu}}()}]{Yuan2023}%
	\BibitemOpen
	\bibfield  {author} {\bibinfo {author} {\bibnamefont {{Zhang, Yanan and Su,
					Dajun and Huang, Yanen and Sun, Hualei and Huo, Mengwu and Shan, Zhaoyang and
					Ye, Kaixin and Yang, Zihan and Li, Rui and Smidman, Michael and Wang, Meng
					and Jiao, Lin and Yuan, Huiqiu}}},\ }\bibfield  {title} {\bibinfo {title}
		{{High-temperature superconductivity with zero-resistance and strange metal
				behavior in $\mathrm{La}_{3}\mathrm{Ni}_{2}\mathrm{O}_{7}$}},\ }\href@noop {}
	{\ }\Eprint {https://arxiv.org/abs/2307.14819} {arXiv:2307.14819}
	\BibitemShut {NoStop}%
	\bibitem [{\citenamefont {Liu}\ \emph {et~al.}()\citenamefont {Liu},
		\citenamefont {Huo}, \citenamefont {Li}, \citenamefont {Li}, \citenamefont
		{Liu}, \citenamefont {Dai}, \citenamefont {Zhou}, \citenamefont {Hao},
		\citenamefont {Lu}, \citenamefont {Wang},\ and\ \citenamefont
		{Wen}}]{WHH2023}%
	\BibitemOpen
	\bibfield  {author} {\bibinfo {author} {\bibfnamefont {Z.}~\bibnamefont
			{Liu}}, \bibinfo {author} {\bibfnamefont {M.}~\bibnamefont {Huo}}, \bibinfo
		{author} {\bibfnamefont {J.}~\bibnamefont {Li}}, \bibinfo {author}
		{\bibfnamefont {Q.}~\bibnamefont {Li}}, \bibinfo {author} {\bibfnamefont
			{Y.}~\bibnamefont {Liu}}, \bibinfo {author} {\bibfnamefont {Y.}~\bibnamefont
			{Dai}}, \bibinfo {author} {\bibfnamefont {X.}~\bibnamefont {Zhou}}, \bibinfo
		{author} {\bibfnamefont {J.}~\bibnamefont {Hao}}, \bibinfo {author}
		{\bibfnamefont {Y.}~\bibnamefont {Lu}}, \bibinfo {author} {\bibfnamefont
			{M.}~\bibnamefont {Wang}},\ and\ \bibinfo {author} {\bibfnamefont {H.-H.}\
			\bibnamefont {Wen}},\ }\bibfield  {title} {\bibinfo {title} {{Electronic
				correlations and energy gap in the bilayer nickelate
				$\mathrm{La}_3\mathrm{Ni}_2\mathrm{O}_7$}},\ }\href@noop {} {\ }\Eprint
	{https://arxiv.org/abs/2307.02950} {arXiv:2307.02950} \BibitemShut {NoStop}%
	\bibitem [{\citenamefont {Liu}\ \emph {et~al.}(2023{\natexlab{a}})\citenamefont
		{Liu}, \citenamefont {Sun}, \citenamefont {Huo}, \citenamefont {Ma},
		\citenamefont {Ji}, \citenamefont {Yi}, \citenamefont {Li}, \citenamefont
		{Liu}, \citenamefont {Yu},\ and\ \citenamefont {Zhang}}]{Liu2023}%
	\BibitemOpen
	\bibfield  {author} {\bibinfo {author} {\bibfnamefont {Z.}~\bibnamefont
			{Liu}}, \bibinfo {author} {\bibfnamefont {H.}~\bibnamefont {Sun}}, \bibinfo
		{author} {\bibfnamefont {M.}~\bibnamefont {Huo}}, \bibinfo {author}
		{\bibfnamefont {X.}~\bibnamefont {Ma}}, \bibinfo {author} {\bibfnamefont
			{Y.}~\bibnamefont {Ji}}, \bibinfo {author} {\bibfnamefont {E.}~\bibnamefont
			{Yi}}, \bibinfo {author} {\bibfnamefont {L.}~\bibnamefont {Li}}, \bibinfo
		{author} {\bibfnamefont {H.}~\bibnamefont {Liu}}, \bibinfo {author}
		{\bibfnamefont {J.}~\bibnamefont {Yu}},\ and\ \bibinfo {author}
		{\bibfnamefont {Z.}~\bibnamefont {Zhang}},\ }\bibfield  {title} {\bibinfo
		{title} {{Evidence for charge and spin density waves in single crystals of
				$\mathrm{La}_{3}\mathrm{Ni}_{2}\mathrm{O}_{7}$ and
				$\mathrm{La}_{3}\mathrm{Ni}_{2}\mathrm{O}_{6}$}},\ }\href
	{https://doi.org/10.1007/s11433-022-1962-4} {\bibfield  {journal} {\bibinfo
			{journal} {Sci. China: Phys., Mech. Astron.}\ }\textbf {\bibinfo {volume}
			{66}},\ \bibinfo {pages} {217411} (\bibinfo {year}
		{2023}{\natexlab{a}})}\BibitemShut {NoStop}%
	\bibitem [{\citenamefont {Chen}\ \emph {et~al.}({\natexlab{b}})\citenamefont
		{Chen}, \citenamefont {Choi}, \citenamefont {Jiang}, \citenamefont {Mei},
		\citenamefont {Jiang}, \citenamefont {Li}, \citenamefont {Agrestini},
		\citenamefont {Garcia-Fernandez}, \citenamefont {Huang}, \citenamefont {Sun},
		\citenamefont {Shen}, \citenamefont {Wang}, \citenamefont {Hu}, \citenamefont
		{Lu}, \citenamefont {Zhou},\ and\ \citenamefont {Feng}}]{RIXS}%
	\BibitemOpen
	\bibfield  {author} {\bibinfo {author} {\bibfnamefont {X.}~\bibnamefont
			{Chen}}, \bibinfo {author} {\bibfnamefont {J.}~\bibnamefont {Choi}}, \bibinfo
		{author} {\bibfnamefont {Z.}~\bibnamefont {Jiang}}, \bibinfo {author}
		{\bibfnamefont {J.}~\bibnamefont {Mei}}, \bibinfo {author} {\bibfnamefont
			{K.}~\bibnamefont {Jiang}}, \bibinfo {author} {\bibfnamefont
			{J.}~\bibnamefont {Li}}, \bibinfo {author} {\bibfnamefont {S.}~\bibnamefont
			{Agrestini}}, \bibinfo {author} {\bibfnamefont {M.}~\bibnamefont
			{Garcia-Fernandez}}, \bibinfo {author} {\bibfnamefont {X.}~\bibnamefont
			{Huang}}, \bibinfo {author} {\bibfnamefont {H.}~\bibnamefont {Sun}}, \bibinfo
		{author} {\bibfnamefont {D.}~\bibnamefont {Shen}}, \bibinfo {author}
		{\bibfnamefont {M.}~\bibnamefont {Wang}}, \bibinfo {author} {\bibfnamefont
			{J.}~\bibnamefont {Hu}}, \bibinfo {author} {\bibfnamefont {Y.}~\bibnamefont
			{Lu}}, \bibinfo {author} {\bibfnamefont {K.-J.}\ \bibnamefont {Zhou}},\ and\
		\bibinfo {author} {\bibfnamefont {D.}~\bibnamefont {Feng}},\ }\bibfield
	{title} {\bibinfo {title} {{Electronic and magnetic excitations in
				$\mathrm{La}_3\mathrm{Ni}_2\mathrm{O}_7$}},\ }\href@noop {} {\
		({\natexlab{b}})},\ \Eprint {https://arxiv.org/abs/2401.12657}
	{arXiv:2401.12657} \BibitemShut {NoStop}%
	\bibitem [{\citenamefont {Dan}\ \emph {et~al.}()\citenamefont {Dan},
		\citenamefont {Zhou}, \citenamefont {Huo}, \citenamefont {Wang},
		\citenamefont {Nie}, \citenamefont {Wang}, \citenamefont {Wu},\ and\
		\citenamefont {Chen}}]{NMR2024}%
	\BibitemOpen
	\bibfield  {author} {\bibinfo {author} {\bibfnamefont {Z.}~\bibnamefont
			{Dan}}, \bibinfo {author} {\bibfnamefont {Y.}~\bibnamefont {Zhou}}, \bibinfo
		{author} {\bibfnamefont {M.}~\bibnamefont {Huo}}, \bibinfo {author}
		{\bibfnamefont {Y.}~\bibnamefont {Wang}}, \bibinfo {author} {\bibfnamefont
			{L.}~\bibnamefont {Nie}}, \bibinfo {author} {\bibfnamefont {M.}~\bibnamefont
			{Wang}}, \bibinfo {author} {\bibfnamefont {T.}~\bibnamefont {Wu}},\ and\
		\bibinfo {author} {\bibfnamefont {X.}~\bibnamefont {Chen}},\ }\href@noop {}
	{\bibinfo {title} {{Spin-density-wave transition in double-layer nickelate
				$\mathrm{La}_3\mathrm{Ni}_2\mathrm{O}_7$}}},\ \Eprint
	{https://arxiv.org/abs/2402.03952} {arXiv:2402.03952} \BibitemShut {NoStop}%
	\bibitem [{\citenamefont {Xie}\ \emph {et~al.}()\citenamefont {Xie},
		\citenamefont {Huo}, \citenamefont {Ni}, \citenamefont {Shen}, \citenamefont
		{Huang}, \citenamefont {Sun}, \citenamefont {Walker}, \citenamefont {Adroja},
		\citenamefont {Yu}, \citenamefont {Shen}, \citenamefont {He}, \citenamefont
		{Cao},\ and\ \citenamefont {Wang}}]{NPD2024}%
	\BibitemOpen
	\bibfield  {author} {\bibinfo {author} {\bibfnamefont {T.}~\bibnamefont
			{Xie}}, \bibinfo {author} {\bibfnamefont {M.}~\bibnamefont {Huo}}, \bibinfo
		{author} {\bibfnamefont {X.}~\bibnamefont {Ni}}, \bibinfo {author}
		{\bibfnamefont {F.}~\bibnamefont {Shen}}, \bibinfo {author} {\bibfnamefont
			{X.}~\bibnamefont {Huang}}, \bibinfo {author} {\bibfnamefont
			{H.}~\bibnamefont {Sun}}, \bibinfo {author} {\bibfnamefont {H.~C.}\
			\bibnamefont {Walker}}, \bibinfo {author} {\bibfnamefont {D.}~\bibnamefont
			{Adroja}}, \bibinfo {author} {\bibfnamefont {D.}~\bibnamefont {Yu}}, \bibinfo
		{author} {\bibfnamefont {B.}~\bibnamefont {Shen}}, \bibinfo {author}
		{\bibfnamefont {L.}~\bibnamefont {He}}, \bibinfo {author} {\bibfnamefont
			{K.}~\bibnamefont {Cao}},\ and\ \bibinfo {author} {\bibfnamefont
			{M.}~\bibnamefont {Wang}},\ }\bibfield  {title} {\bibinfo {title} {{Neutron
				Scattering Studies on the High-$T_c$ Superconductor
				$\mathrm{La}_3\mathrm{Ni}_2\mathrm{O}_{7-\delta}$ at Ambient Pressure}},\
	}\href@noop {} {\ }\Eprint {https://arxiv.org/abs/2401.12635}
	{arXiv:2401.12635} \BibitemShut {NoStop}%
	\bibitem [{\citenamefont {Niedermayer}\ \emph {et~al.}(1998)\citenamefont
		{Niedermayer}, \citenamefont {Bernhard}, \citenamefont {Blasius},
		\citenamefont {Golnik}, \citenamefont {Moodenbaugh},\ and\ \citenamefont
		{Budnick}}]{Niedermayer1998}%
	\BibitemOpen
	\bibfield  {author} {\bibinfo {author} {\bibfnamefont {C.}~\bibnamefont
			{Niedermayer}}, \bibinfo {author} {\bibfnamefont {C.}~\bibnamefont
			{Bernhard}}, \bibinfo {author} {\bibfnamefont {T.}~\bibnamefont {Blasius}},
		\bibinfo {author} {\bibfnamefont {A.}~\bibnamefont {Golnik}}, \bibinfo
		{author} {\bibfnamefont {A.}~\bibnamefont {Moodenbaugh}},\ and\ \bibinfo
		{author} {\bibfnamefont {J.~I.}\ \bibnamefont {Budnick}},\ }\bibfield
	{title} {\bibinfo {title} {{Common Phase Diagram for Antiferromagnetism in
				${\mathrm{La}}_{2\ensuremath{-}\mathit{x}}{\mathrm{Sr}}_{\mathit{x}}{\mathrm{CuO}}_{4}$
				and
				${Y}_{1\ensuremath{-}\mathit{x}}{\mathrm{Ca}}_{\mathit{x}}{\mathrm{Ba}}_{2}{\mathrm{Cu}}_{3}{O}_{6}$
				as Seen by Muon Spin Rotation}},\ }\href
	{https://doi.org/10.1103/PhysRevLett.80.3843} {\bibfield  {journal} {\bibinfo
			{journal} {Phys. Rev. Lett.}\ }\textbf {\bibinfo {volume} {80}},\ \bibinfo
		{pages} {3843} (\bibinfo {year} {1998})}\BibitemShut {NoStop}%
	\bibitem [{\citenamefont {Julien}(2003)}]{Julien2003}%
	\BibitemOpen
	\bibfield  {author} {\bibinfo {author} {\bibfnamefont {M.-H.}\ \bibnamefont
			{Julien}},\ }\bibfield  {title} {\bibinfo {title} {{Magnetic order and
				superconductivity in
				$\mathrm{La}_{2-\ensuremath{x}}\mathrm{Sr}_{\mathrm{x}}\mathrm{CuO}_4$: a
				review}},\ }\href
	{https://doi.org/https://doi.org/10.1016/S0921-4526(02)01997-X} {\bibfield
		{journal} {\bibinfo  {journal} {Phys. B (Amsterdam, Neth.)}\ }\textbf
		{\bibinfo {volume} {329-333}},\ \bibinfo {pages} {693} (\bibinfo {year}
		{2003})},\ \bibinfo {note} {proceedings of the 23rd International Conference
		on Low Temperature Physics}\BibitemShut {NoStop}%
	\bibitem [{\citenamefont {Sanna}\ \emph {et~al.}(2004)\citenamefont {Sanna},
		\citenamefont {Allodi}, \citenamefont {Concas}, \citenamefont {Hillier},\
		and\ \citenamefont {Renzi}}]{Sanna2004}%
	\BibitemOpen
	\bibfield  {author} {\bibinfo {author} {\bibfnamefont {S.}~\bibnamefont
			{Sanna}}, \bibinfo {author} {\bibfnamefont {G.}~\bibnamefont {Allodi}},
		\bibinfo {author} {\bibfnamefont {G.}~\bibnamefont {Concas}}, \bibinfo
		{author} {\bibfnamefont {A.~D.}\ \bibnamefont {Hillier}},\ and\ \bibinfo
		{author} {\bibfnamefont {R.~D.}\ \bibnamefont {Renzi}},\ }\bibfield  {title}
	{\bibinfo {title} {{Nanoscopic Coexistence of Magnetism and Superconductivity
				in
				${\mathrm{Y}\mathrm{B}\mathrm{a}}_{2}{\mathrm{C}\mathrm{u}}_{3}{\mathrm{O}}_{6+x}$
				Detected by Muon Spin Rotation}},\ }\href
	{https://doi.org/10.1103/PhysRevLett.93.207001} {\bibfield  {journal}
		{\bibinfo  {journal} {Phys. Rev. Lett.}\ }\textbf {\bibinfo {volume} {93}},\
		\bibinfo {pages} {207001} (\bibinfo {year} {2004})}\BibitemShut {NoStop}%
	\bibitem [{\citenamefont {Paglione}\ and\ \citenamefont
		{Greene}(2010)}]{Paglione2010}%
	\BibitemOpen
	\bibfield  {author} {\bibinfo {author} {\bibfnamefont {J.}~\bibnamefont
			{Paglione}}\ and\ \bibinfo {author} {\bibfnamefont {R.~L.}\ \bibnamefont
			{Greene}},\ }\bibfield  {title} {\bibinfo {title} {High-temperature
			superconductivity in iron-based materials},\ }\href
	{https://doi.org/10.1038/nphys1759} {\bibfield  {journal} {\bibinfo
			{journal} {Nat. Phys.}\ }\textbf {\bibinfo {volume} {6}},\ \bibinfo {pages}
		{645} (\bibinfo {year} {2010})}\BibitemShut {NoStop}%
	\bibitem [{\citenamefont {Dai}(2015)}]{Dai2015}%
	\BibitemOpen
	\bibfield  {author} {\bibinfo {author} {\bibfnamefont {P.}~\bibnamefont
			{Dai}},\ }\bibfield  {title} {\bibinfo {title} {Antiferromagnetic order and
			spin dynamics in iron-based superconductors},\ }\href
	{https://doi.org/10.1103/RevModPhys.87.855} {\bibfield  {journal} {\bibinfo
			{journal} {Rev. Mod. Phys.}\ }\textbf {\bibinfo {volume} {87}},\ \bibinfo
		{pages} {855} (\bibinfo {year} {2015})}\BibitemShut {NoStop}%
	\bibitem [{\citenamefont {Kenzelmann}\ \emph {et~al.}(2008)\citenamefont
		{Kenzelmann}, \citenamefont {Strässle}, \citenamefont {Niedermayer},
		\citenamefont {Sigrist}, \citenamefont {Padmanabhan}, \citenamefont
		{Zolliker}, \citenamefont {Bianchi}, \citenamefont {Movshovich},
		\citenamefont {Bauer}, \citenamefont {Sarrao},\ and\ \citenamefont
		{Thompson}}]{Kenzelmann2008}%
	\BibitemOpen
	\bibfield  {author} {\bibinfo {author} {\bibfnamefont {M.}~\bibnamefont
			{Kenzelmann}}, \bibinfo {author} {\bibfnamefont {T.}~\bibnamefont
			{Strässle}}, \bibinfo {author} {\bibfnamefont {C.}~\bibnamefont
			{Niedermayer}}, \bibinfo {author} {\bibfnamefont {M.}~\bibnamefont
			{Sigrist}}, \bibinfo {author} {\bibfnamefont {B.}~\bibnamefont
			{Padmanabhan}}, \bibinfo {author} {\bibfnamefont {M.}~\bibnamefont
			{Zolliker}}, \bibinfo {author} {\bibfnamefont {A.~D.}\ \bibnamefont
			{Bianchi}}, \bibinfo {author} {\bibfnamefont {R.}~\bibnamefont {Movshovich}},
		\bibinfo {author} {\bibfnamefont {E.~D.}\ \bibnamefont {Bauer}}, \bibinfo
		{author} {\bibfnamefont {J.~L.}\ \bibnamefont {Sarrao}},\ and\ \bibinfo
		{author} {\bibfnamefont {J.~D.}\ \bibnamefont {Thompson}},\ }\bibfield
	{title} {\bibinfo {title} {{Coupled Superconducting and Magnetic Order in
				$\mathrm{CeCoIn}_5$}},\ }\href {https://doi.org/10.1126/science.1161818}
	{\bibfield  {journal} {\bibinfo  {journal} {Science}\ }\textbf {\bibinfo
			{volume} {321}},\ \bibinfo {pages} {1652} (\bibinfo {year}
		{2008})}\BibitemShut {NoStop}%
	\bibitem [{\citenamefont {Blundell}(1999)}]{Blundell1999}%
	\BibitemOpen
	\bibfield  {author} {\bibinfo {author} {\bibfnamefont {S.~J.}\ \bibnamefont
			{Blundell}},\ }\bibfield  {title} {\bibinfo {title} {{Spin-polarized muons in
				condensed matter physics}},\ }\href {https://doi.org/10.1080/001075199181521}
	{\bibfield  {journal} {\bibinfo  {journal} {Contemp. Phys.}\ }\textbf
		{\bibinfo {volume} {40}},\ \bibinfo {pages} {175} (\bibinfo {year}
		{1999})}\BibitemShut {NoStop}%
			\bibitem [{\citenamefont {{A. Yaouanc and P. Dalmas de
			R{\'e}otier}}(2011)}]{Yaouanc2011MuonSR}%
	\BibitemOpen
	\bibfield  {author} {\bibinfo {author} {\bibnamefont {{A. Yaouanc and P.
				Dalmas de R{\'e}otier}}},\ }\href@noop {} {\textit{\bibinfo {title}
		{{\textit{Muon Spin Rotation, Relaxation, and Resonance: Applications to
					Condensed Matter}}}}}\ (\bibinfo  {publisher} {Oxford University Press,
	Oxford, UK},\ \bibinfo {year} {2011})\BibitemShut {NoStop}%
	\bibitem [{Adr(2022)}]{Adrian2022}%
	\BibitemOpen
	\bibfield  {title} {\bibinfo {title} {{Muon spin spectroscopy}},\ }\href
	{https://doi.org/10.1038/s43586-022-00094-x} {\bibfield  {journal} {\bibinfo
			{journal} {Nat. Rev. Methods Primers}\ }\textbf {\bibinfo {volume} {2}},\
		\bibinfo {pages} {5} (\bibinfo {year} {2022})}\BibitemShut {NoStop}%
	\bibitem [{\citenamefont {Rodriguez-Carvajal}(1990)}]{FULLPROF}%
	\BibitemOpen
	\bibfield  {author} {\bibinfo {author} {\bibfnamefont {J.}~\bibnamefont
			{Rodriguez-Carvajal}},\ }\bibfield  {title} {\bibinfo {title} {{FULLPROF: a
				program for Rietveld refinement and pattern matching analysis}},\ }in\
	\href@noop {} {\textit {\bibinfo {booktitle} {satellite meeting on powder
				diffraction of the XV congress of the IUCr}}},\ Vol.\ \bibinfo {volume}
	{127}\ (\bibinfo {organization} {Toulouse, France},\ \bibinfo {year}
	{1990})\BibitemShut {NoStop}%
	\bibitem [{\citenamefont {Suter}\ and\ \citenamefont {Wojek}(2012)}]{MUSRFIT}%
	\BibitemOpen
	\bibfield  {author} {\bibinfo {author} {\bibfnamefont {A.}~\bibnamefont
			{Suter}}\ and\ \bibinfo {author} {\bibfnamefont {B.}~\bibnamefont {Wojek}},\
	}\bibfield  {title} {\bibinfo {title} {{Musrfit: A Free Platform-Independent
				Framework for $\mu$SR Data Analysis}},\ }\href
	{https://doi.org/https://doi.org/10.1016/j.phpro.2012.04.042} {\bibfield
		{journal} {\bibinfo  {journal} {Phys. Procedia}\ }\textbf {\bibinfo {volume}
			{30}},\ \bibinfo {pages} {69} (\bibinfo {year} {2012})}\BibitemShut {NoStop}%
	\bibitem [{\citenamefont {Zhang}\ \emph {et~al.}(1994)\citenamefont {Zhang},
		\citenamefont {Greenblatt},\ and\ \citenamefont {Goodenough}}]{Zhang1994}%
	\BibitemOpen
	\bibfield  {author} {\bibinfo {author} {\bibfnamefont {Z.}~\bibnamefont
			{Zhang}}, \bibinfo {author} {\bibfnamefont {M.}~\bibnamefont {Greenblatt}},\
		and\ \bibinfo {author} {\bibfnamefont {J.~B.}\ \bibnamefont {Goodenough}},\
	}\bibfield  {title} {\bibinfo {title} {{Synthesis, Structure, and Properties
				of the Layered Perovskite
				$\mathrm{La}_{3}\mathrm{Ni}_{2}\mathrm{O}_{7-\delta}$}},\ }\href
	{https://doi.org/https://doi.org/10.1006/jssc.1994.1059} {\bibfield
		{journal} {\bibinfo  {journal} {J. Solid State Chem.}\ }\textbf {\bibinfo
			{volume} {108}},\ \bibinfo {pages} {402} (\bibinfo {year}
		{1994})}\BibitemShut {NoStop}%
	\bibitem [{\citenamefont {Liu}\ \emph {et~al.}(2023{\natexlab{b}})\citenamefont
		{Liu}, \citenamefont {Mei}, \citenamefont {Ye}, \citenamefont {Chen},\ and\
		\citenamefont {Yang}}]{Liu_PRL2023}%
	\BibitemOpen
	\bibfield  {author} {\bibinfo {author} {\bibfnamefont {Y.-B.}\ \bibnamefont
			{Liu}}, \bibinfo {author} {\bibfnamefont {J.-W.}\ \bibnamefont {Mei}},
		\bibinfo {author} {\bibfnamefont {F.}~\bibnamefont {Ye}}, \bibinfo {author}
		{\bibfnamefont {W.-Q.}\ \bibnamefont {Chen}},\ and\ \bibinfo {author}
		{\bibfnamefont {F.}~\bibnamefont {Yang}},\ }\bibfield  {title} {\bibinfo
		{title} {{${\mathrm{s}}^{\ifmmode\pm\else\textpm\fi{}}$-Wave Pairing and the
				Destructive Role of Apical-Oxygen Deficiencies in
				${\mathrm{La}}_{3}{\mathrm{Ni}}_{2}{\mathrm{O}}_{7}$ under Pressure}},\
	}\href {https://doi.org/10.1103/PhysRevLett.131.236002} {\bibfield  {journal}
		{\bibinfo  {journal} {Phys. Rev. Lett.}\ }\textbf {\bibinfo {volume} {131}},\
		\bibinfo {pages} {236002} (\bibinfo {year} {2023}{\natexlab{b}})}\BibitemShut
	{NoStop}%
	\bibitem [{\citenamefont {Hayano}\ \emph {et~al.}(1979)\citenamefont {Hayano},
		\citenamefont {Uemura}, \citenamefont {Imazato}, \citenamefont {Nishida},
		\citenamefont {Yamazaki},\ and\ \citenamefont {Kubo}}]{Hayano1979}%
	\BibitemOpen
	\bibfield  {author} {\bibinfo {author} {\bibfnamefont {R.~S.}\ \bibnamefont
			{Hayano}}, \bibinfo {author} {\bibfnamefont {Y.~J.}\ \bibnamefont {Uemura}},
		\bibinfo {author} {\bibfnamefont {J.}~\bibnamefont {Imazato}}, \bibinfo
		{author} {\bibfnamefont {N.}~\bibnamefont {Nishida}}, \bibinfo {author}
		{\bibfnamefont {T.}~\bibnamefont {Yamazaki}},\ and\ \bibinfo {author}
		{\bibfnamefont {R.}~\bibnamefont {Kubo}},\ }\bibfield  {title} {\bibinfo
		{title} {Zero-and low-field spin relaxation studied by positive muons},\
	}\href {https://doi.org/10.1103/PhysRevB.20.850} {\bibfield  {journal}
		{\bibinfo  {journal} {Phys. Rev. B}\ }\textbf {\bibinfo {volume} {20}},\
		\bibinfo {pages} {850} (\bibinfo {year} {1979})}\BibitemShut {NoStop}%
	\bibitem [{\citenamefont {Amato}(1997)}]{Amato1997}%
	\BibitemOpen
	\bibfield  {author} {\bibinfo {author} {\bibfnamefont {A.}~\bibnamefont
			{Amato}},\ }\bibfield  {title} {\bibinfo {title} {{Heavy-fermion systems
				studied by \ensuremath{\mu}SR technique}},\ }\href
	{https://doi.org/10.1103/RevModPhys.69.1119} {\bibfield  {journal} {\bibinfo
			{journal} {Rev. Mod. Phys.}\ }\textbf {\bibinfo {volume} {69}},\ \bibinfo
		{pages} {1119} (\bibinfo {year} {1997})}\BibitemShut {NoStop}%
	\bibitem [{\citenamefont {Zhang}\ \emph {et~al.}(2015)\citenamefont {Zhang},
		\citenamefont {MacLaughlin}, \citenamefont {Hillier}, \citenamefont {Ding},
		\citenamefont {Huang}, \citenamefont {Maple},\ and\ \citenamefont
		{Shu}}]{Zhang2015}%
	\BibitemOpen
	\bibfield  {author} {\bibinfo {author} {\bibfnamefont {J.}~\bibnamefont
			{Zhang}}, \bibinfo {author} {\bibfnamefont {D.~E.}\ \bibnamefont
			{MacLaughlin}}, \bibinfo {author} {\bibfnamefont {A.~D.}\ \bibnamefont
			{Hillier}}, \bibinfo {author} {\bibfnamefont {Z.~F.}\ \bibnamefont {Ding}},
		\bibinfo {author} {\bibfnamefont {K.}~\bibnamefont {Huang}}, \bibinfo
		{author} {\bibfnamefont {M.~B.}\ \bibnamefont {Maple}},\ and\ \bibinfo
		{author} {\bibfnamefont {L.}~\bibnamefont {Shu}},\ }\bibfield  {title}
	{\bibinfo {title} {{Broken time-reversal symmetry in superconducting
				${\mathrm{Pr}}_{1\ensuremath{-}x}{\mathrm{Ce}}_{x}{\mathrm{Pt}}_{4}{\mathrm{Ge}}_{12}$}},\
	}\href {https://doi.org/10.1103/PhysRevB.91.104523} {\bibfield  {journal}
		{\bibinfo  {journal} {Phys. Rev. B}\ }\textbf {\bibinfo {volume} {91}},\
		\bibinfo {pages} {104523} (\bibinfo {year} {2015})}\BibitemShut {NoStop}%
	\bibitem [{\citenamefont {Sugiyama}\ \emph {et~al.}(2009)\citenamefont
		{Sugiyama}, \citenamefont {Månsson}, \citenamefont {Ikedo}, \citenamefont
		{Goko}, \citenamefont {Mukai}, \citenamefont {Andreica}, \citenamefont
		{Amato}, \citenamefont {Ariyoshi},\ and\ \citenamefont
		{Ohzuku}}]{Sugiyama2009}%
	\BibitemOpen
	\bibfield  {author} {\bibinfo {author} {\bibfnamefont {J.}~\bibnamefont
			{Sugiyama}}, \bibinfo {author} {\bibfnamefont {M.}~\bibnamefont {Månsson}},
		\bibinfo {author} {\bibfnamefont {Y.}~\bibnamefont {Ikedo}}, \bibinfo
		{author} {\bibfnamefont {T.}~\bibnamefont {Goko}}, \bibinfo {author}
		{\bibfnamefont {K.}~\bibnamefont {Mukai}}, \bibinfo {author} {\bibfnamefont
			{D.}~\bibnamefont {Andreica}}, \bibinfo {author} {\bibfnamefont
			{A.}~\bibnamefont {Amato}}, \bibinfo {author} {\bibfnamefont
			{K.}~\bibnamefont {Ariyoshi}},\ and\ \bibinfo {author} {\bibfnamefont
			{T.}~\bibnamefont {Ohzuku}},\ }\bibfield  {title} {\bibinfo {title}
		{{${\ensuremath{\mu}}^{+}\text{SR}$ investigation of local magnetic order in
				${\text{LiCrO}}_{2}$}},\ }\href {https://doi.org/10.1103/PhysRevB.79.184411}
	{\bibfield  {journal} {\bibinfo  {journal} {Phys. Rev. B}\ }\textbf {\bibinfo
			{volume} {79}},\ \bibinfo {pages} {184411} (\bibinfo {year}
		{2009})}\BibitemShut {NoStop}%
	\bibitem [{Sup()}]{Supmat}%
	\BibitemOpen
	\href@noop {} {\bibinfo {title} {See supplemental material at}},\ \bibinfo
	{howpublished} {"\url{URL_will_be_inserted_by_publisher}"}\BibitemShut
	{NoStop}%
	\bibitem [{\citenamefont {Campostrini}\ \emph {et~al.}(2002)\citenamefont
		{Campostrini}, \citenamefont {Hasenbusch}, \citenamefont {Pelissetto},
		\citenamefont {Rossi},\ and\ \citenamefont {Vicari}}]{Critical_exponent}%
	\BibitemOpen
	\bibfield  {author} {\bibinfo {author} {\bibfnamefont {M.}~\bibnamefont
			{Campostrini}}, \bibinfo {author} {\bibfnamefont {M.}~\bibnamefont
			{Hasenbusch}}, \bibinfo {author} {\bibfnamefont {A.}~\bibnamefont
			{Pelissetto}}, \bibinfo {author} {\bibfnamefont {P.}~\bibnamefont {Rossi}},\
		and\ \bibinfo {author} {\bibfnamefont {E.}~\bibnamefont {Vicari}},\
	}\bibfield  {title} {\bibinfo {title} {{Critical exponents and equation of
				state of the three-dimensional Heisenberg universality class}},\ }\href
	{https://doi.org/10.1103/PhysRevB.65.144520} {\bibfield  {journal} {\bibinfo
			{journal} {Phys. Rev. B}\ }\textbf {\bibinfo {volume} {65}},\ \bibinfo
		{pages} {144520} (\bibinfo {year} {2002})}\BibitemShut {NoStop}%
	\bibitem [{\citenamefont {Bramwell}\ and\ \citenamefont
		{Holdsworth}(1993)}]{Bramwell_1993}%
	\BibitemOpen
	\bibfield  {author} {\bibinfo {author} {\bibfnamefont {S.~T.}\ \bibnamefont
			{Bramwell}}\ and\ \bibinfo {author} {\bibfnamefont {P.~C.~W.}\ \bibnamefont
			{Holdsworth}},\ }\bibfield  {title} {\bibinfo {title} {{Magnetization and
				universal sub-critical behaviour in two-dimensional XY magnets}},\ }\href
	{https://doi.org/10.1088/0953-8984/5/4/004} {\bibfield  {journal} {\bibinfo
			{journal} {Journal of Physics: Condensed Matter}\ }\textbf {\bibinfo {volume}
			{5}},\ \bibinfo {pages} {L53} (\bibinfo {year} {1993})}\BibitemShut {NoStop}%
	\bibitem [{\citenamefont {Huddart}\ \emph {et~al.}(2022)\citenamefont
		{Huddart}, \citenamefont {Hernández-Melián}, \citenamefont {Hicken},
		\citenamefont {Gomilšek}, \citenamefont {Hawkhead}, \citenamefont {Clark},
		\citenamefont {Pratt},\ and\ \citenamefont {Lancaster}}]{MuFinder}%
	\BibitemOpen
	\bibfield  {author} {\bibinfo {author} {\bibfnamefont {B.}~\bibnamefont
			{Huddart}}, \bibinfo {author} {\bibfnamefont {A.}~\bibnamefont
			{Hernández-Melián}}, \bibinfo {author} {\bibfnamefont {T.}~\bibnamefont
			{Hicken}}, \bibinfo {author} {\bibfnamefont {M.}~\bibnamefont {Gomilšek}},
		\bibinfo {author} {\bibfnamefont {Z.}~\bibnamefont {Hawkhead}}, \bibinfo
		{author} {\bibfnamefont {S.}~\bibnamefont {Clark}}, \bibinfo {author}
		{\bibfnamefont {F.}~\bibnamefont {Pratt}},\ and\ \bibinfo {author}
		{\bibfnamefont {T.}~\bibnamefont {Lancaster}},\ }\bibfield  {title} {\bibinfo
		{title} {{MuFinder: A program to determine and analyse muon stopping
				sites}},\ }\href {https://doi.org/https://doi.org/10.1016/j.cpc.2022.108488}
	{\bibfield  {journal} {\bibinfo  {journal} {Comput. Phys. Commun.}\ }\textbf
		{\bibinfo {volume} {280}},\ \bibinfo {pages} {108488} (\bibinfo {year}
		{2022})}\BibitemShut {NoStop}%
	\bibitem [{\citenamefont {Clark}\ \emph {et~al.}(2005)\citenamefont {Clark},
		\citenamefont {Segall}, \citenamefont {Pickard}, \citenamefont {Hasnip},
		\citenamefont {Probert}, \citenamefont {Refson},\ and\ \citenamefont
		{Payne}}]{Castep}%
	\BibitemOpen
	\bibfield  {author} {\bibinfo {author} {\bibfnamefont {S.~J.}\ \bibnamefont
			{Clark}}, \bibinfo {author} {\bibfnamefont {M.~D.}\ \bibnamefont {Segall}},
		\bibinfo {author} {\bibfnamefont {C.~J.}\ \bibnamefont {Pickard}}, \bibinfo
		{author} {\bibfnamefont {P.~J.}\ \bibnamefont {Hasnip}}, \bibinfo {author}
		{\bibfnamefont {M.~I.~J.}\ \bibnamefont {Probert}}, \bibinfo {author}
		{\bibfnamefont {K.}~\bibnamefont {Refson}},\ and\ \bibinfo {author}
		{\bibfnamefont {M.~C.}\ \bibnamefont {Payne}},\ }\bibfield  {title} {\bibinfo
		{title} {{First principles methods using CASTEP}},\ }\href
	{https://doi.org/doi:10.1524/zkri.220.5.567.65075} {\bibfield  {journal}
		{\bibinfo  {journal} {Z. fur Krist. - Cryst. Mater.}\ }\textbf {\bibinfo
			{volume} {220}},\ \bibinfo {pages} {567} (\bibinfo {year}
		{2005})}\BibitemShut {NoStop}%
	\bibitem [{\citenamefont {Pratt}(2019)}]{DipoleCal}%
	\BibitemOpen
	\bibfield  {author} {\bibinfo {author} {\bibfnamefont {F.~L.}\ \bibnamefont
			{Pratt}},\ }\bibfield  {title} {\bibinfo {title} {Dipolar field calculations
			for muon spectroscopy}\ }\href {https://doi.org/10.5281/zenodo.3476167}
	{10.5281/zenodo.3476167} (\bibinfo {year} {2019})\BibitemShut {NoStop}%
	\bibitem [{\citenamefont {Khasanov}\ \emph {et~al.}()\citenamefont {Khasanov},
		\citenamefont {Hicken}, \citenamefont {Gawryluk}, \citenamefont {Sorel},
		\citenamefont {Bötzel}, \citenamefont {Lechermann}, \citenamefont {Eremin},
		\citenamefont {Luetkens},\ and\ \citenamefont {Guguchia}}]{Zurab}%
	\BibitemOpen
	\bibfield  {author} {\bibinfo {author} {\bibfnamefont {R.}~\bibnamefont
			{Khasanov}}, \bibinfo {author} {\bibfnamefont {T.~J.}\ \bibnamefont
			{Hicken}}, \bibinfo {author} {\bibfnamefont {D.~J.}\ \bibnamefont
			{Gawryluk}}, \bibinfo {author} {\bibfnamefont {L.~P.}\ \bibnamefont {Sorel}},
		\bibinfo {author} {\bibfnamefont {S.}~\bibnamefont {Bötzel}}, \bibinfo
		{author} {\bibfnamefont {F.}~\bibnamefont {Lechermann}}, \bibinfo {author}
		{\bibfnamefont {I.~M.}\ \bibnamefont {Eremin}}, \bibinfo {author}
		{\bibfnamefont {H.}~\bibnamefont {Luetkens}},\ and\ \bibinfo {author}
		{\bibfnamefont {Z.}~\bibnamefont {Guguchia}},\ }\bibfield  {title} {\bibinfo
		{title} {{Pressure-Induced Split of the Density Wave Transitions in
				$\mathrm{La}_3\mathrm{Ni}_2\mathrm{O}_{7-\delta}$}},\ }\href@noop {} {\
	}\Eprint {https://arxiv.org/abs/2402.10485} {arXiv:2402.10485} \BibitemShut
	{NoStop}%
	\bibitem [{\citenamefont {Bonilla}\ \emph {et~al.}(2011)\citenamefont
		{Bonilla}, \citenamefont {Marcano}, \citenamefont {Herrero-Albillos},
		\citenamefont {Maisuradze}, \citenamefont {Garc\'{\i}a},\ and\ \citenamefont
		{Bartolom\'e}}]{Bonilla2011}%
	\BibitemOpen
	\bibfield  {author} {\bibinfo {author} {\bibfnamefont {C.~M.}\ \bibnamefont
			{Bonilla}}, \bibinfo {author} {\bibfnamefont {N.}~\bibnamefont {Marcano}},
		\bibinfo {author} {\bibfnamefont {J.}~\bibnamefont {Herrero-Albillos}},
		\bibinfo {author} {\bibfnamefont {A.}~\bibnamefont {Maisuradze}}, \bibinfo
		{author} {\bibfnamefont {L.~M.}\ \bibnamefont {Garc\'{\i}a}},\ and\ \bibinfo
		{author} {\bibfnamefont {F.}~\bibnamefont {Bartolom\'e}},\ }\bibfield
	{title} {\bibinfo {title} {{$\ensuremath{\mu}$SR study of short-range
				magnetic order in the paramagnetic regime of ErCo${}_{2}$}},\ }\href
	{https://doi.org/10.1103/PhysRevB.84.184425} {\bibfield  {journal} {\bibinfo
			{journal} {Phys. Rev. B}\ }\textbf {\bibinfo {volume} {84}},\ \bibinfo
		{pages} {184425} (\bibinfo {year} {2011})}\BibitemShut {NoStop}%
	\bibitem [{\citenamefont {Taniguchi}\ \emph {et~al.}(1995)\citenamefont
		{Taniguchi}, \citenamefont {Nishikawa}, \citenamefont {Yasui}, \citenamefont
		{Kobayashi}, \citenamefont {Takeda}, \citenamefont {Shamoto},\ and\
		\citenamefont {Sato}}]{Taniguchi1995}%
	\BibitemOpen
	\bibfield  {author} {\bibinfo {author} {\bibfnamefont {S.}~\bibnamefont
			{Taniguchi}}, \bibinfo {author} {\bibfnamefont {T.}~\bibnamefont
			{Nishikawa}}, \bibinfo {author} {\bibfnamefont {Y.}~\bibnamefont {Yasui}},
		\bibinfo {author} {\bibfnamefont {Y.}~\bibnamefont {Kobayashi}}, \bibinfo
		{author} {\bibfnamefont {J.}~\bibnamefont {Takeda}}, \bibinfo {author}
		{\bibfnamefont {S.-i.}\ \bibnamefont {Shamoto}},\ and\ \bibinfo {author}
		{\bibfnamefont {M.}~\bibnamefont {Sato}},\ }\bibfield  {title} {\bibinfo
		{title} {{Transport, Magnetic and Thermal Properties of
				$\mathrm{La}_{3}\mathrm{Ni}_{2}\mathrm{O}_{7-\delta}$}},\ }\href
	{https://doi.org/10.1143/JPSJ.64.1644} {\bibfield  {journal} {\bibinfo
			{journal} {J. Phys. Soc. Jpn.}\ }\textbf {\bibinfo {volume} {64}},\ \bibinfo
		{pages} {1644} (\bibinfo {year} {1995})}\BibitemShut {NoStop}%
	\bibitem [{\citenamefont {Wu}\ \emph {et~al.}(2001)\citenamefont {Wu},
		\citenamefont {Neumeier},\ and\ \citenamefont {Hundley}}]{Wu2001}%
	\BibitemOpen
	\bibfield  {author} {\bibinfo {author} {\bibfnamefont {G.}~\bibnamefont
			{Wu}}, \bibinfo {author} {\bibfnamefont {J.~J.}\ \bibnamefont {Neumeier}},\
		and\ \bibinfo {author} {\bibfnamefont {M.~F.}\ \bibnamefont {Hundley}},\
	}\bibfield  {title} {\bibinfo {title} {{Magnetic susceptibility, heat
				capacity, and pressure dependence of the electrical resistivity of
				${\mathrm{La}}_{3}{\mathrm{Ni}}_{2}{\mathrm{O}}_{7}$ and
				${\mathrm{La}}_{4}{\mathrm{Ni}}_{3}{\mathrm{O}}_{10}$}},\ }\href
	{https://doi.org/10.1103/PhysRevB.63.245120} {\bibfield  {journal} {\bibinfo
			{journal} {Phys. Rev. B}\ }\textbf {\bibinfo {volume} {63}},\ \bibinfo
		{pages} {245120} (\bibinfo {year} {2001})}\BibitemShut {NoStop}%
	\bibitem [{\citenamefont {Dong}\ \emph {et~al.}()\citenamefont {Dong},
		\citenamefont {Huo}, \citenamefont {Li}, \citenamefont {Li}, \citenamefont
		{Li}, \citenamefont {Sun}, \citenamefont {Lu}, \citenamefont {Wang},
		\citenamefont {Wang},\ and\ \citenamefont {Chen}}]{MEP}%
	\BibitemOpen
	\bibfield  {author} {\bibinfo {author} {\bibfnamefont {Z.}~\bibnamefont
			{Dong}}, \bibinfo {author} {\bibfnamefont {M.}~\bibnamefont {Huo}}, \bibinfo
		{author} {\bibfnamefont {J.}~\bibnamefont {Li}}, \bibinfo {author}
		{\bibfnamefont {J.}~\bibnamefont {Li}}, \bibinfo {author} {\bibfnamefont
			{P.}~\bibnamefont {Li}}, \bibinfo {author} {\bibfnamefont {H.}~\bibnamefont
			{Sun}}, \bibinfo {author} {\bibfnamefont {Y.}~\bibnamefont {Lu}}, \bibinfo
		{author} {\bibfnamefont {M.}~\bibnamefont {Wang}}, \bibinfo {author}
		{\bibfnamefont {Y.}~\bibnamefont {Wang}},\ and\ \bibinfo {author}
		{\bibfnamefont {Z.}~\bibnamefont {Chen}},\ }\bibfield  {title} {\bibinfo
		{title} {{Visualization of Oxygen Vacancies and Self-doped Ligand Holes in
				$\mathrm{La}_3\mathrm{Ni}_2\mathrm{O}_{7-\delta}$}},\ }\href@noop {} {\
	}\Eprint {https://arxiv.org/abs/2312.15727} {arXiv:2312.15727} \BibitemShut
	{NoStop}%
\end{thebibliography}
	\end{document}